\documentclass[aps, prl, superscriptaddress, showpacs, floatfix, 10pt, twocolumn, reprint]{revtex4-2}

\usepackage{graphicx}
\usepackage{xcolor}
\usepackage{transparent}
\usepackage{import}
\usepackage{overpic}
\usepackage{amssymb}
\usepackage{mathtools}
\usepackage{bbm}
\usepackage{bm}
\usepackage{mathrsfs}
\usepackage{verbatim}
\usepackage{dsfont}
\usepackage{calligra}
\usepackage{amsthm}
\usepackage{booktabs}
\usepackage{multirow}
\usepackage{nicematrix}
\usepackage{comment}

\usepackage[breaklinks=true,colorlinks,citecolor=blue,linkcolor=blue,urlcolor=blue]{hyperref}

\makeatletter
\DeclareRobustCommand{\cev}[1]{%
  {\mathpalette\do@cev{#1}}%
}
\newcommand{\do@cev}[2]{%
  \vbox{\offinterlineskip
    \sbox\z@{$\m@th#1 x$}%
    \ialign{##\cr
      \hidewidth\reflectbox{$\m@th#1\vec{}\mkern4mu$}\hidewidth\cr
      \noalign{\kern-\ht\z@}
      $\m@th#1#2$\cr
    }%
  }%
}
\makeatother

\newcommand{\tr}{\operatorname{Tr}}
\newcommand{\hc}{{\rm H.c.}}
\newcommand{\1}{\mathds{1}}
\renewcommand{\Re}{\operatorname{Re}}
\renewcommand{\Im}{\operatorname{Im}}
\newcommand{\clb}[1]{{\color{blue} #1}}

\renewcommand{\a}[1]{a^{\vphantom{\dagger}}_{#1}}
\newcommand{\ad}[1]{a^{\dagger}_{#1}}
\renewcommand{\b}[1]{b^{\vphantom{\dagger}}_{#1}}
\newcommand{\bd}[1]{b^{\dagger}_{#1}}

\renewcommand{\H}[1]{{\cal H}_{#1}}

\newcommand{\G}{{\cal G}}

\newcommand{\D}{{\cal D}}
\newcommand{\C}{{\cal C}}

\newcommand{\M}{{\cal M}}
\renewcommand{\L}{{\cal L}}
\newcommand{\N}{{\cal N}}
\renewcommand{\P}{{\cal P}}

\renewcommand{\AA}{{\mathcal A}}
\newcommand{\J}{{\mathcal J}}
\newcommand{\V}{{\cal V}}

\newcommand{\U}{{\cal U}}
\newcommand{\Z}{{\cal Z}}
\newcommand{\I}{{\cal I}}

\renewcommand{\O}{{\cal O}}

\renewcommand{\d}{{\rm d}}

\newcommand{\vev}[1]{\bigl \langle #1 \bigr \rangle_0}
\newcommand{\tord}{{\mathbb{T}}}
\newcommand{\drot}{{\tilde{\mathbb{T}}}}

\newcommand{\arcoth}{\operatorname{arcoth}}
\newcommand{\sectiontitle}[1]{\clb{\it #1.}}
\newcommand{\sech}{\operatorname{sech}}

\newif\ifappendix
\appendixtrue 

\begin{document}

\title{Controlling the non-Markovianity of quantum Brownian motion}

\date{\today}

\author{Guglielmo Pellitteri}\email{guglielmo.pellitteri@sns.it}
\affiliation{Scuola Normale Superiore, 56126 Pisa, Italy}

\author{Vittorio Giovannetti}
\affiliation{Scuola Normale Superiore, 56126 Pisa, Italy}
\affiliation{NEST and Istituto Nanoscienze-CNR, 56126 Pisa, Italy}

\author{Vasco Cavina}
\affiliation{Scuola Normale Superiore, 56126 Pisa, Italy} 

\begin{abstract}
    We analyze the exact dynamics of a generalized quantum Brownian motion model, employing Gaussian master equation methods. We demonstrate that, by modulating the relative weights of specific interaction channels, we can control the degree of non-Markovianity of the system, and induce a transition from non-Markovian to Markovian regimes. The non-Markovianity of the evolution is formally characterized by leveraging the Gorini--Kossakowski--Sudarshan--Lindblad theorem and by employing quantitative measures of information backflow. Finally, we clarify the physical mechanism behind the phenomenology of this model, thereby providing a systematic platform for environment engineering through the strategic tuning of dissipation channels.
\end{abstract}

\maketitle

\sectiontitle{Introduction}---Dissipative quantum systems in the strong-coupling regime are fundamentally non-Markovian, exhibiting persistent memory effects that invalidate the standard memoryless approximation~\cite{Li2020Perspective, deVega2017review, Breuer2016Review, Chruscinski2022}. Elucidating the nature of memory effects is thus a central challenge in the study of open quantum systems, leading to the development of diverse formalisms for their characterization~\cite{BLPpaper,rivas2010markovian, vacchini2011markovianity, pezzutto2026nonclassical}. Significant theoretical effort has been devoted to developing robust methods for characterizing the dynamics of non-Markovian quantum systems, ranging from stochastic approaches~\cite{stockburger2001diffusion, stockburger2002exact, stockburger2004simulating, tanimura2006review, tilloy2017unraveling, zhou2005stochastic, gasbarri2018stochastic, yan2018stochastic, cavina2025unifying} to perturbative expansions~\cite{grifoni1998review, colla2025unveiling, colla2025recursive} and non-equilibrium many-body methods~\cite{Talarico2019scalable}.

For Gaussian systems, a more rigorous and analytically precise characterization is possible. This is achieved by exploiting the mathematical properties of Gaussian completely positive maps~\cite{heinosaari2009semigroup, Torreetal2015, LiuzzoScorpoetal2017, holevo2012quantum} in conjunction with exact master equation formalisms~\cite{ferialdi2016master, Hu1992Classic, Zhang2012ME, d2025exact,  pellitteri2026exact}, which provide a complete description of the dynamics beyond the Markovian limit.

The possibility of controlling the degree of non-Markovianity has been theorized in several physical systems~\cite{Lorenzoetal2012, Tangetal2023, roccati2024controlling}, and experimentally realized in diverse setups, ranging from trapped ion systems~\cite{Wittemer2018}, spin qubits in diamond-based nitrogen-vacancy centers~\cite{Haaseetal2018}, molecular spin~\cite{Kukitaetal2020} and photonic setups~\cite{Liu2011Experimental, Tang2012Measuring, Liu2013Photonic, Cialdi2011Programmable, Cialdi2017AllOptical, Chiuri2012Linear, Jin2015AllOptical}. This has provided a better understanding of the role of memory effects in different settings and has improved the ability of manipulating open quantum system dynamics, with the aim of counteracting the deleterious impact of environmental noise.

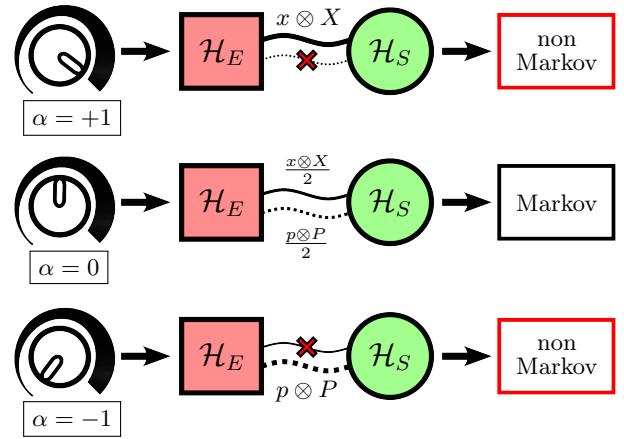
\begin{figure}
    \centering
    \begin{overpic}[width = 0.98\linewidth]{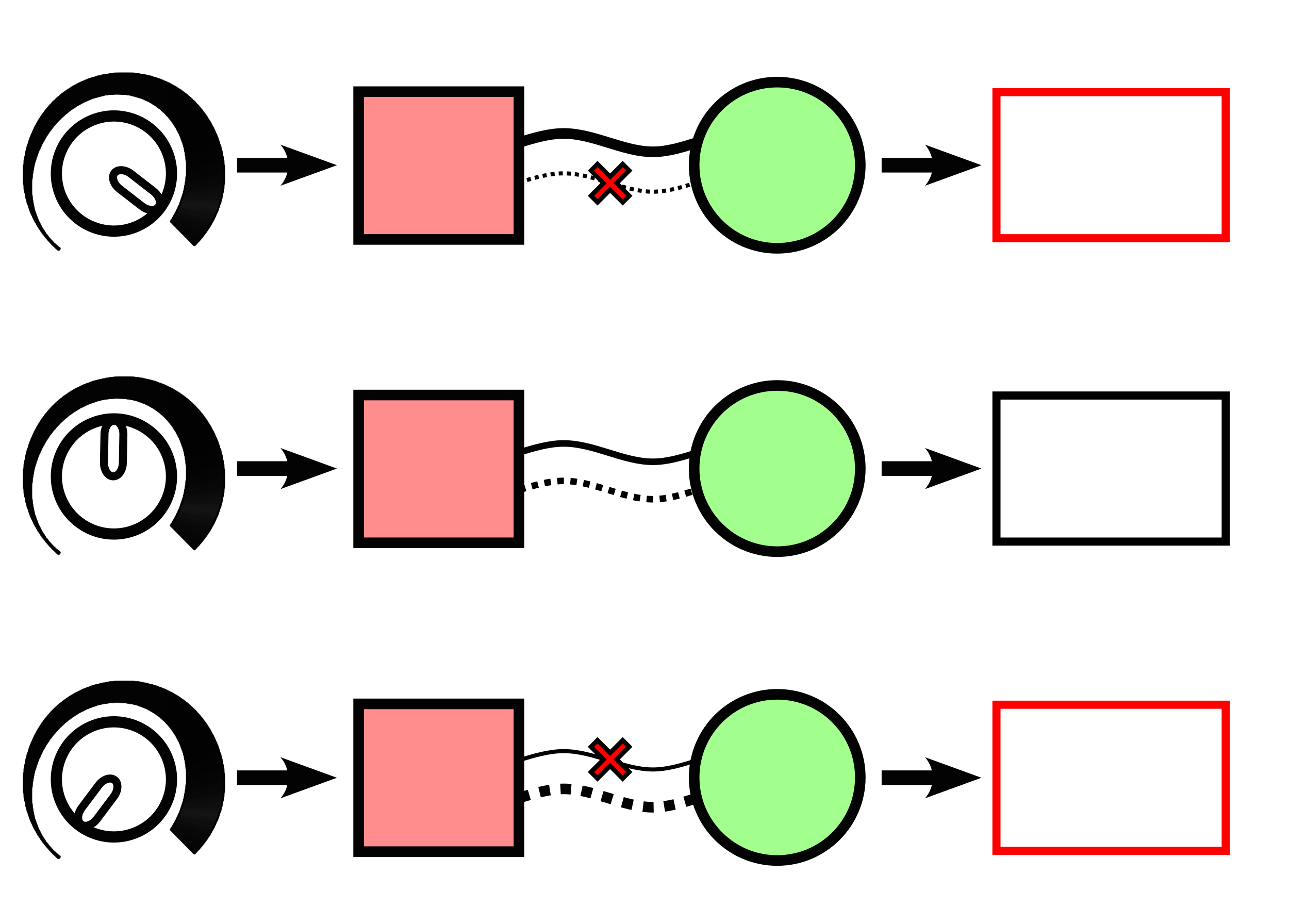}
        \put(3, 47.5){$\boxed{\alpha = +1}$}
        \put(4, 24.){$\boxed{\alpha = 0}$}
        \put(3, 0.5){$\boxed{\alpha = -1}$}
        \put(30.5, 57.5){\large $\H{E}$}
        \put(57, 57.5){\large$\H{S}$}
        \put(30.5, 34.){\large$\H{E}$}
        \put(57, 34.){\large$\H{S}$}
        \put(30.5, 10.){\large$\H{E}$}
        \put(57, 10.){\large$\H{S}$}
        \put(42.5, 63.){$x\otimes X$} 
        \put(43.5, 39.5){$\frac{x\otimes X}{2}$}
        \put(43.5, 28.){$\frac{p\otimes P}{2}$}
        \put(42.5, 5.){$p\otimes P$}
        \put(80, 56){\shortstack{non \\ Markov}}
        \put(80, 34){Markov}
        \put(80, 8.5){\shortstack{non \\ Markov}}
    \end{overpic}
    \caption{Schematic representation of the generalized quantum Brownian motion model employed in this work for different values of the control parameter $\alpha$. For $\alpha = \pm 1$ system and environment interact in the $x \otimes X$ ($p \otimes P$) channel only, number conservation symmetry is broken, and the resulting evolution is non-Markovian. For $\alpha = 0$, the spectral weight of the two channels is balanced, the total number of excitations is conserved, and the resulting evolution is Markovian.}
    \label{fig:1}
\end{figure}

In this work, we study the exact dynamics of a generalized quantum Brownian motion (gQBM) model, employing the machinery of exact Gaussian master equations~\cite{d2025exact}. The gQBM model is a generalization of the standard Caldeira--Leggett model~\cite{caldeira1983path, leggett1987dynamics, einsiedler2020nonmarkov, Ferialdi2017} which allows for tunable interaction in different channels, i.e. coupling through different operator products. We prove that, by tuning the relative weight of such channels, the degree of non-Markovianity of the evolution can be precisely controlled, allowing both Markovian and non-Markovian regimes within the same framework, as pictured in Fig.~\ref{fig:1}. 

This work is organized as follows. First, we describe the gQBM model and give its exact solution in the form of a Gaussian master equation. Second, we give a formal definition of non-Markovianity as CP-divisibility, and provide a necessary and sufficient condition for CP-divisibility by leveraging the Gorini--Kossakowski--Sudarshan--Lindblad (GKSL) theorem. We then derive a general quasi-GKSL form for Gaussian master equations, which enables us to apply the CP-divisibility condition to Gaussian models. Applying this construction to the gQBM model, we show that we can control the non-Markovianity of the latter by tuning the relative weight of different interaction channels. Finally, we quantify the degree of non-Markovianity of the model as a function of such relative weight by employing quantitative measures of information backflow, and we draw our conclusions.

\sectiontitle{Model and exact solution}---In the gQBM model, a system is bilinearly coupled to a collection of quadratic modes (the environment) through two interaction mechanisms, one associating the oscillator's position with the position of the modes and the other involving their momenta. The Hamiltonian reads
\begin{equation}
\label{eq:total-hamiltonian}
    \H{} = \H{S} + \H{E} + \V_\alpha~,
\end{equation}
where $\mathcal{H}_S$, $\mathcal{H}_E$, $\mathcal{V}_{\alpha}$ are the system, environment, and interaction Hamiltonians, respectively defined as
\begin{align}
\label{eq:system-hamiltonian}
    \H{S}  &= \frac{{p}^2}{2m} + \frac{1}{2}m\omega_0^2  x^2~,\\
\label{eq:environment-hamiltonian}
    \H{E}  &= \sum_{n} \left(\frac{p_n^2}{2M_n} + \frac{1}{2} M_n \omega_n^2 x_n^2\right)~,\\
\label{eq:interaction-hamiltonian}
    \V_\alpha  &=\frac{1+\alpha}{2}{x} \otimes {X} + \frac{1-\alpha}{2} p \otimes P~,
\end{align}
where $x, p$ are the system quadratures, $x_n, p_n$ are the quadratures relative to the $n$-th environmental mode, and $X := \sum_{n}f_n x_n$, $P:= \sum_n g_n p_n$, with $f_n, g_n$ coupling constants. Finally, $\alpha$ is a dimensionless parameter which controls the relative weight of the two interaction channels. In the present work, we consider the environment to be prepared in a thermal state at inverse temperature $\beta$. In the following, for the sake of clarity, we fix $\hbar = 1$, as well as $m = \omega_0 = 1$. 

Since both the system and the environment Hamiltonians are quadratic in the coordinates, we can describe the exact dynamics of the gQBM model via a Gaussian master equation (GME)~\cite{d2025exact}, given by
\begin{equation}
\label{eq:gme}
\begin{split}
   & \dot \rho_S(t) +i\bigl[{\H{}}_S,  \rho_S(t)\bigr]\\
   &=\sum_{i,j}\int_0^t\d \tau\,\G^>_{ij}(t, \tau)\,\bigl[ r_j(\tau-t)  \rho_S(t), r_i\bigr] + \hc~,
\end{split}
\end{equation}
where $\bm r = (x, p)^T$ is the quadrature vector, $\bm r(s):= e^{i\H{S}s} \bm r(0)e^{-i\H{S}s}$ evolves according to the free system Hamiltonian, and $\G^>(t,\tau)$ is a kernel accounting for the correlations established between system and environment between times $\tau$ and $t$. This kernel contains all the information about the influence of $E$ on $S$ and is fully determined by the temperature of the former, as well as by the environmental operators appearing in the coupling Hamiltonian~\eqref{eq:interaction-hamiltonian}.
Eq.~\eqref{eq:gme} remains valid even in the regime of arbitrarily strong system-environment coupling and fully captures memory effects. 
In the weak-coupling limit, Eq.~\eqref{eq:gme} reduces to the standard Redfield equation~\cite{breuer2007book, deVega2017review}, with which it shares the operatorial form.
Further details on the derivation and solution of Eq.~\eqref{eq:gme} and to its application to the gQBM model are provided in App.~\ref{app:gme}. 

\sectiontitle{Non-Markovianity and CP-divisibility}---
From a formal point of view, the dynamics of an open quantum system evolving between two times $\tau$, $t$ can be described by a family of linear, completely positive (CP) and trace-preserving maps $\{\Phi_{t,\tau}, \,t\geq \tau \}$.
In this framework, non-Markovianity is typically characterized according to the Rivas--Huelga--Plenio (RHP) criterion~\cite{rivas2010entanglement}, which associates Markovianity with CP-divisibility of the evolution~\cite{Rivasetal2014, vacchini2011markovianity, Breuer2016Review}. The latter is the property of $\Phi_{t, \tau}$ of satisfying the following composition rule:
\begin{equation}
    \Phi_{t, \tau} = \Phi_{t, s} \circ \Phi_{s, \tau}~,\quad \forall s \in [\tau, t]~. \label{eq:composition}
\end{equation}
The physical intuition behind this requirement is that, using dilation techniques, the two maps in the r.h.s. of Eq.~\eqref{eq:composition} can be seen as independent processes induced by the subsequent contact with separate, uncorrelated environments.

\begin{figure}
    \centering
    \begin{overpic}[width = 1.\linewidth]{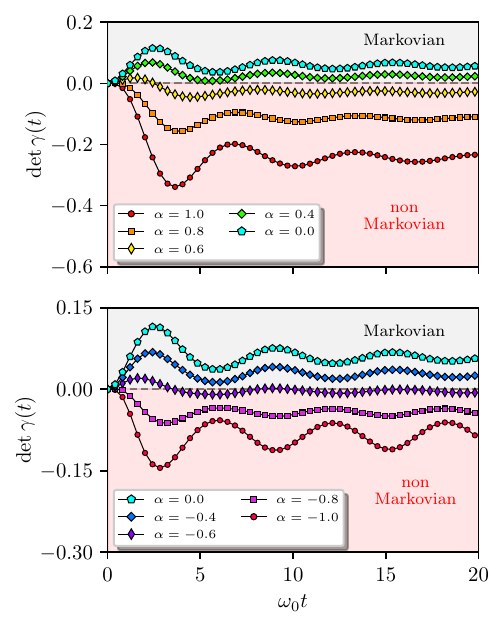}
        \put(2,96){(a)}
        \put(2,50){(b)}
        \put(28.5, 47){{\footnotesize $\alpha = 0$}}
        \put(28.5, 24.5){{\footnotesize$\alpha = -1$}}
        \put(28.5, 92){{\footnotesize $\alpha = 0$}}
        \put(31., 69.5){{\footnotesize$\alpha = +1$}}
    \end{overpic}
    \caption{Behavior of the determinant of the Kossakowski matrix $\gamma(t)$ as a function of time, for different values of the control parameter $\alpha$. (a) $\alpha \in [0, 1]$. (b) $\alpha \in [-1, 0]$. The condition $\det \gamma(t)\geq 0$ is necessary and sufficient for Markovianity of the evolution. For details on the numerics and parameter values we refer to App.~\ref{app:parameters}.}
    \label{fig:2}
\end{figure}

\begin{figure}[t]
    \centering
    \begin{overpic}[width = 1.\linewidth]{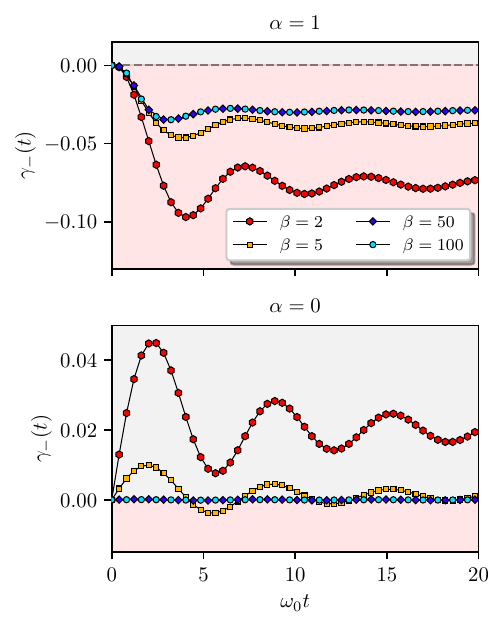}
        \put(2,96){(a)}
        \put(2,50){(b)}
        \put(45, 15.){$\gamma_- \simeq 0$ as $\beta \to \infty$}
        \put(45, 85.5){$\gamma_- < 0$ as $\beta \to \infty$}
    \end{overpic}
    \caption{Behavior of the lesser canonical rate $\gamma_-(t)$ as a function of time for increasing values of the inverse temperature of environment, $\beta$. (a) For $\alpha = 1$, $\gamma_-(t)$ remains finite and negative as $\beta$ approaches infinity, due to processes allowed by the presence of CRTs. (b) For $\alpha = 0$, in the absence of CRTs, $\gamma_-(t) \to 0$ as $\beta \to \infty$, so the dynamics becomes fully Markovian in the high-temperature regime.}
    \label{fig:3}
\end{figure}
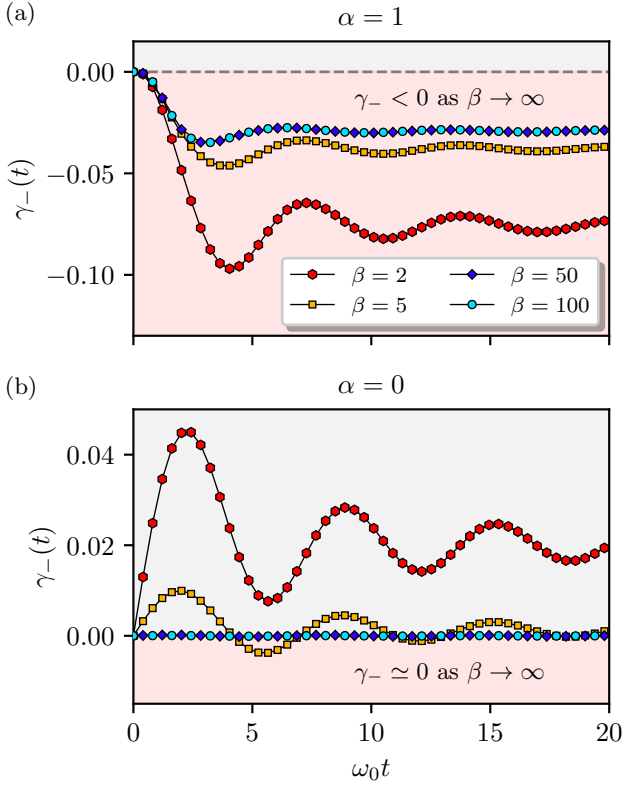

In the case in which the generator of $\Phi_{t, \tau}$ exists, we formally introduce it as
\begin{equation}
    \L(t) \coloneqq \left[\frac{\d}{\d t} \Phi_{t, \tau}^{\vphantom{-1}} \right]\Phi^{-1}_{t, \tau}~,
\end{equation}
and we can express the CP-divisibility requirement as a condition on $\mathcal{L}(t)$, rather than on the map itself. The Gorini--Kossakowski--Sudarshan--Lindblad (GKSL) theorem \cite{gorini1976classic, lindblad1976classic}, holding for Gaussian semigroups $\Phi_{t, \tau}$~\cite{holevo2012quantum, heinosaari2009semigroup}, states that the generators of a CP-divisible map can be written in the quasi-GKSL form~\cite{quasiGKSL_note, gorini1976classic, lindblad1976classic, Megieretal2017, hall2014canonical}
\begin{equation}
\label{eq:dissipator}
    \mathcal{L}(t)\bullet{}
   = - i \bigl[{\Lambda{}}(t), {}\bullet{}\bigr] + \sum_{i, j} \gamma_{ij}(t) \D_{ij}(t) {}\bullet{}~,
\end{equation}
where $\Lambda(t)$ is a self-adjoint operator, $\gamma(t) = \gamma^\dagger(t) \geq 0$ is the so-called Kossakowski matrix and $\D_{ij}(t)$  are superoperators of the form
\begin{equation}
\label{eq:general-dissipator}
    \D_{ij}(t) {}\bullet{} \coloneqq L_i(t){} \bullet{} L_j^\dagger(t) -\frac{1}{2} \bigl\{L_j^\dagger(t) L_i(t), {}\bullet{}\bigr\}~.
\end{equation}
Since we can always write the generator of a CP Gaussian map in quasi-GKSL form~\cite{heinosaari2009semigroup}, the property of CP-divisibility is equivalent to global positive-semidefiniteness of the Kossakowski 
matrix $\gamma(t)$, i.e.
\begin{equation}
\label{eq:cp-criterion}
    \Phi_{t,\tau}\text{ is CP-divisible} \iff \gamma(s) \geq 0 \quad\forall s \in [\tau, t]~.
\end{equation}
This condition is particularly convenient for the present analysis, as it will be sufficient to derive a quasi-GKSL representation of Eq.~\eqref{eq:gme} and check the RHP criterion~\eqref{eq:cp-criterion}.

\sectiontitle{Controlling non-Markovianity}---
The results presented here~[Eqs.~\eqref{eq:gamma}-\eqref{eq:M-matrix}] hold in general for any Gaussian open evolution whose dissipator is defined by a GME, i.e.
\begin{equation}
\label{eq:gme-dissipator}
    \begin{split}
    &\L(t){}\bullet{} + i \bigl[\H{S}, {}\bullet{} \bigr] \\
    & = \sum_{i,j}\int_0^t\d \tau\,\G^>_{ij}(t, \tau)\,\bigl[ A_j(\tau-t)  \bullet{}, A_i^\dagger(0)\bigr] + \hc ~,
    \end{split}
\end{equation}
with $\{A_i(s)\}_i$ interaction-picture system operators. Eq.~\eqref{eq:gme} is an example of dissipator that falls within this class. 
Our goal is to recast Eq.~\eqref{eq:gme-dissipator} in the form given in Eq.~\eqref{eq:dissipator}, by formulating $\gamma(t)$, $\Lambda(t)$ and $\mathcal{D}_{ij}(t)$ in terms of the memory kernel $\mathcal{G}_{ij}^{>}(t, \tau)$.
Under the Gaussian hypothesis, we can write any system operator $A_i$ as a linear combination of quadratures, i.e. $A_i(s) := \sum_j\AA_{ij}(s) r_j$. This leads, after some manipulations, to an equation in quasi-GKSL form~\eqref{eq:dissipator} (see App.~\ref{app:gksl} for the complete derivation), with
\begin{align}
\label{eq:gamma}
    \gamma(t) &= \M^*(t)  + \M^T(t) ~,\\
\label{eq:Lambda}
    \Lambda(t)  & = \H{S}+  \frac{i}{2}\,\bm r^T \bigl[\M^*(t) - \M^T(t) \bigr] \bm r~,\\
\label{eq:dissipator2}
    \D_{ij}\bullet{}& := r_i \bullet{} r_j - \frac{1}{2}\bigl\{r_j r_i, \bullet{}\bigr\}~,
\end{align}
where $^*$ denotes complex conjugation, $^T$ denotes matrix transposition and we defined
\begin{equation}
\label{eq:M-matrix}
    \M(t):= \int_0^t\d \tau\, \AA^\dagger(0)\G^>(t, \tau)\AA(\tau-t)~.
\end{equation}
%
Applying this general construction to the gQBM dynamics given by Eq.~\eqref{eq:gme} allows us to characterize its non-Markovian features via the RHP criterion~\eqref{eq:cp-criterion}. In the following, we will assume that the coupling constants $f_n$ and $g_n$ [see below Eq.~\eqref{eq:interaction-hamiltonian}] are chosen in such a way that $f_n = M_n \omega_n g_n$ $\forall n$. In this way, the environment is characterized by a single density of states, which is chosen to be sub-ohmic with an exponential cutoff, as specified in App.~\ref{app:parameters}. Thus, the relative weight of the $x \otimes X$ and $p \otimes P$ coupling channels is uniquely quantified by the dimensionless parameter $\alpha$.  

We start by presenting the case in which the environment is prepared in a Gibbs state with fixed inverse temperature $\beta = 1/2\omega_0 = 1/2$.
Since $\tr \gamma(t) \geq 0$ for all $t$ [see App.~\ref{app:parameters}], the RHP criterion is equivalent to the condition $\det \gamma(t) \geq 0$.
As illustrated in Fig.~\ref{fig:2}, the control parameter $\alpha$ acts as a dial for memory effects. For $\alpha = \pm 1$, the system exhibits {\it eternal} non-Markovianity~\cite{Megieretal2017, hall2014canonical}, where the lesser of the two canonical rates $\gamma_{\pm}(t)$---i.e. the eigenvalues of $\gamma(t)$---remains strictly negative for all $t > 0$. In these regimes, our model recovers the physics of the standard Caldeira--Leggett model~\cite{caldeira1983path, leggett1987dynamics, breuer2007book, einsiedler2020nonmarkov}, consistently with known results~\cite{Ferialdi2017}. Conversely, as $\alpha$ approaches zero, the dynamics become strictly Markovian ($\gamma(t) \geq 0$ $\forall t$), demonstrating that $\alpha$ provides a direct mechanism to tune the emergence of quantum memory. While the dissipator associated with $\gamma_+(t)$ consistently acts as a standard decoherence channel, the $\gamma_-(t)$ channel is functionally flexible: it can be engineered to switch from driving decoherence to inducing recoherence, effectively reversing the flow of information back into the system.

The transition from Markovian to eternally non-Markovian dynamics in the gQBM can be explained by symmetry considerations. To obtain insight on such transition, a weak-coupling analysis of the Kossakowski matrix in the $t \to \infty$ limit is sufficient. By rewriting the interaction Hamiltonian in Eq.~\eqref{eq:interaction-hamiltonian} in terms of creation and annihilation operators, we obtain
\begin{equation}
\label{eq:interaction-hamiltonian2}
    \V_{\alpha} = \ad{} B +\a{} B^\dagger + \alpha\bigl(\a{} B +\ad{} B^\dagger \bigr) ~,
\end{equation}
where $B:= \sum_n \sqrt{f_n g_n/2} \,b_n$ is a linear combination of annihilation operators of the environment. 
For $\alpha=0$, the total number of excitations $N = \ad{}\a{}+ \sum_n \bd{n}\b{n}$ is conserved, 
while this is not the case for  $\alpha \neq 0$ due to the presence of counter-rotating terms (CRTs). Thus, in the absence of CRTs, choosing $\a{}$ and $\ad{}$ as Lindblad operators $\{L_i\}_i$ in the dissipator~\eqref{eq:general-dissipator} and imposing the number conservation constraint, we see that the Kossakowski matrix $\gamma^\infty := \lim_{t\to \infty} \gamma(t)$ is diagonal in this basis and positive-semidefinite, as proven analytically in the weak-coupling limit [see App.~\ref{app:crt}] and numerically in the exact case.

Dissipators that do not conserve the total excitation number introduce off-diagonal terms in $\gamma^\infty$ (in the $\a{}, \ad{}$ basis) and allow to violate positive-semidefiniteness of the latter. Via weak-coupling analysis, we determine that the condition
\begin{equation}
    \alpha > \sech \left(\frac{\beta \omega_0}{2}\right)
\end{equation}
is sufficient for $\gamma^\infty \ngeq 0$, and thus for eternally non-Markovian dynamics.
We note that a connection between the breakdown of non-Markovianity and the rotating-wave approximation, which consists in the elimination of CRTs on the basis of timescale separation arguments, was already observed in Ref.~\cite{Makela2013}.

To gain further intuition on the behavior of decoherence in this model, we investigate its low-temperature limit ($\beta \to \infty$). For $\alpha \neq 0$, as illustrated in Fig.~\ref{fig:3}a, the presence of CRTs allows for processes which give rise to a recoherence rate $\gamma_-(t) < 0$ $\forall t > 0$ which is finite and negative even in the zero-temperature limit. Thus, when the environment is arbitrarily close to its ground state, vacuum fluctuations mediated by CRTs are still capable of driving a persistent information backflow to the system. Conversely, when $\alpha = 0$, we have $\gamma_{-}(t) \to 0$ for $\beta \to \infty$ [see Fig.~\ref{fig:3}b], and the low-temperature dynamics is approximately described by the following Markovian master equation:
\begin{equation}
\label{eq:lowT-master}
    \dot \rho_S  + i \bigl[\Lambda(t), \rho_S\bigr]\simeq \gamma_{+}(t) \Bigl[\a{}\rho_S \ad{} - \frac{1}{2}\bigl\{\ad{} \a{}, \rho_S\bigr\}\Bigr]~,
\end{equation}
with $\gamma_+(t)\geq 0$ $\forall t\geq0$. Equation~\eqref{eq:lowT-master} shows that, for $\alpha = 0$ and at zero temperature, the structured bath degrades into a memoryless pure dissipation sink that solely absorbs excitations via the annihilation operator $\a{}$. The full derivation of this limit is given in App.~\ref{app:lowT}. In the high-temperature limit $\beta \to 0$, memory effects are suppressed due to unbounded thermal fluctuations, as expected from the standard Caldeira--Leggett master equation formalism~\cite{caldeira1983path, leggett1987dynamics}. 

\begin{figure}
    \centering
    \begin{overpic}[width=1.\linewidth]{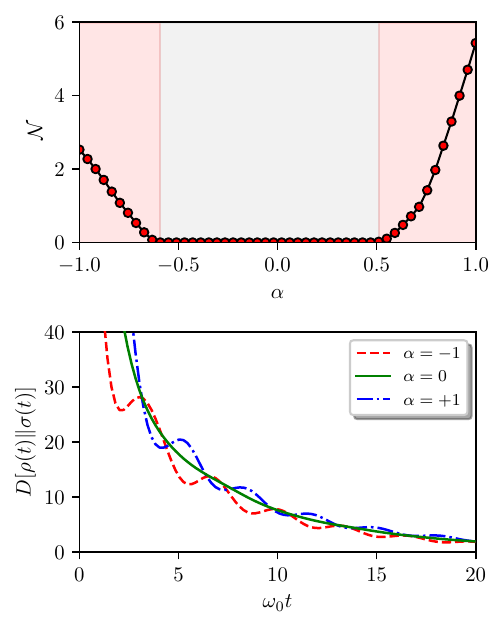}
        \put(2, 96){(a)}
        \put(2, 48){(b)}
        \put(15, 90){$\gamma \ngeq 0$}
        \put(64, 90){$\gamma \ngeq 0$}
        \put(40, 90){$\gamma \geq 0$}
    \end{overpic}
    \caption{(a) BLP measure of non-Markovianity based on contractivity of the Umegaki relative entropy, as a function of the control parameter $\alpha \in[-1, 1]$. In the regions marked in red, the Kossakowski matrix is not positive-semidefinite, and we have $\N > 0$. In the gray region, we have $\gamma(t)\geq 0$ $\forall t$ and therefore $\N = 0$. (b) Umegaki relative entropy between two differently prepared states as a function of time for $\alpha = 0, \pm 1$. Local non-contractivity of $D[\rho(t)\| \sigma(t)]$ is due to information backflow and thus a witness of non-Markovianity. In both (a) and (b), initial states $\rho(0),\sigma(0)$ are quasi-coherent states with displacements $\bar{\bm r}_\rho = - \bar{ \bm r}_\sigma = (3, -3)^T$ and covariance matrices $\sigma_\rho = \sigma_\sigma = (1+\epsilon)\1$, with $\epsilon\ll 1$. Numerical parameters are specified in App.~\ref{app:parameters}, and details on information backflow measure are reported in App.~\ref{app:measures}.}
    \label{fig:4}
\end{figure}
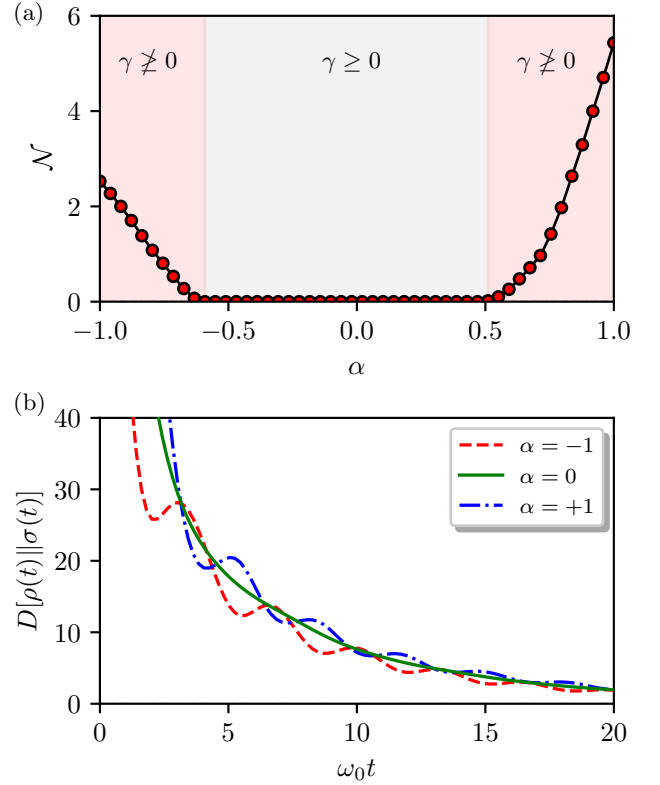

\sectiontitle{Measures of information backflow}---Another way to identify non-Markovian effects on quantum dynamics is the Breuer--Laine--Piilo (BLP) measure~\cite{BLPpaper}. Let us consider a measure for the distinguishability $D[\rho \|\sigma]$ between two quantum states $\rho$, $\sigma$---not necessarily a distance {\it stricto sensu}---obeying the requirement of contractivity under CP maps~\cite{Ruskai1994}, i.e.
\begin{equation}
\label{eq:contractivity}
    D[\Phi\rho \|\Phi\sigma] \leq D[\rho\|\sigma]\quad \forall\rho, \sigma~,
\end{equation}
with $\Phi$ CP. If a quantum channel $\Phi_{t,\tau}$ does not satisfy the property of CP-divisibility, the contractivity property~\eqref{eq:contractivity} can be violated locally in time, implying information backflow from the environment to the system. Therefore, the condition
\begin{equation}
\label{eq:lambda}
    \lambda(t; \rho, \sigma) :=\frac{\d}{\d t'} D[\Phi_{t',\tau}\rho \|\Phi_{t',\tau}\sigma] \bigg|_{t' = t} >0~,
\end{equation}
for some instant $t\geq \tau$ and an arbitrary pair of states $\rho, \sigma$ is a witness of information backflow, and thus a sufficient condition for CP-indivisibility of the channel $\Phi_{t, \tau}$. On the basis of this definition, one can construct a measure of non-Markovianity of the process based on the total increase in distinguishability during the evolution. Such a measure is defined as~\cite{BLPpaper}
\begin{equation}
\label{eq:Nmax}
    \N_{\rm max} := \max_{\rho, \sigma} \int_{\lambda > 0} \d t\,\lambda(t; \rho, \sigma) ~.
\end{equation}
Here, the time integration is restricted to all the intervals in which $\lambda > 0$, and the maximum is taken over all pairs of initial states. 

In the original paper~\cite{BLPpaper}, the suggested measure for distinguishability is the trace distance. Here, we employ the Umegaki relative entropy
\begin{equation}
\label{eq:umegaki}
    D[\rho \| \sigma] := \tr \bigl[\rho \bigl(\log \rho - \log \sigma\bigr)\bigr]~,
\end{equation}
which belongs to a class of entropies which are remarkably easy to calculate for continuous-variable Gaussian states~\cite{Seshadreesan2018}. Following Ref.~\cite{einsiedler2020nonmarkov}, we omit the maximization appearing in Eq.~\eqref{eq:Nmax} and investigate the quantity
\begin{equation}
\label{eq:N}
    \N :=\int_{\lambda > 0} \d t\,\lambda(t; \rho, \sigma) \leq \N_{\rm max}
\end{equation}
for fixed pairs of initial states, as detailed in App.~\ref{app:measures}. This quantity is shown in Fig.~\ref{fig:4}a as a function of the control parameter $\alpha \in[-1, 1]$. Strikingly, in the intervals in which $\N > 0$, we have $\gamma(t) \ngeq 0$ for some or for all instants $t$ on which the integral in Eq.~\eqref{eq:N} is performed, with $\N$ monotonically increasing in $|\alpha|$. Fig.~\ref{fig:4}b showcases local violation of the contractivity property~\eqref{eq:contractivity} of the Umegaki relative entropy due to information backflow under non-Markovian evolution. 

An alternative witness of non-Markovianity, also introduced by Rivas, Huelga and Plenio~\cite{rivas2010entanglement}, is the revival of entanglement between the system and an ancilla. The non-monotonicity of any entanglement measure (such as the logarithmic negativity of entanglement) is a sufficient condition for non-Markovianity of the evolution. Notably, for what concerns the parameter space explored in this work, this witness is not able to detect non-Markovianity, as shown in detail in App.~\ref{app:measures}.

\sectiontitle{Conclusions}---We gave a detailed analysis of non-Markovianity in a gQBM model. This analysis is based on the sufficient and necessary condition for CP-indivisibility provided by the GKSL theorem for continuous-variable Gaussian systems, and corroborated by the BLP witness of information backflow. We showed that tuning the relative weight between the two interaction channels in the gQBM Hamiltonian allows us to control the degree of non-Markovianity of the system, enabling the transition between eternally non-Markovian and fully-Markovian regimes. The physical intuition behind this phenomenology is found in timescale separation arguments connected to the presence of symmetry-breaking terms in the interaction Hamiltonian. 

The possibility of exerting control over memory effects in quantum dynamics could find promising applications in dark-state engineering~\cite{lidar1998decoherence}, quantum error mitigation and correction~\cite{liu2024non, wang2025non} and quantum reservoir computing~\cite{sannia2025non}. Moreover, it paves the way for a clearer comprehension of complex quantum dynamics, and for the study of memory effects on the thermodynamics of open quantum systems~\cite{abiuso2019non, strasberg2016nonequilibrium, thomas2018thermodynamics, zhang2014quantum, pellitteri2026exact}. 

\sectiontitle{Acknowledgments}---We thank Alexander Holevo, Andrea Smirne, Bassano Vacchini, Dario De Santis and Ludovico Lami for useful discussions. We acknowledge financial support by MUR (Ministero dell’Universit{\`a} e della Ricerca) through the PNRR MUR project PE0000023-NQSTI. 

\bibliography{bibliography.bib}

\ifappendix

\appendix
\onecolumngrid

\setcounter{secnumdepth}{2} 
\setcounter{equation}{0}
\setcounter{figure}{0}
\setcounter{table}{0} 

\renewcommand{\thesection}{\Alph{section}}
\renewcommand{\thesubsection}{\thesection.\arabic{subsection}}
\renewcommand{\theequation}{\thesection\arabic{equation}}
\renewcommand{\thefigure}{\thesection\arabic{figure}}
\renewcommand{\theHfigure}{A.\thefigure}
\renewcommand{\theHequation}{A.\theequation}

\section{Exact Gaussian master equation for the gQBM model}
\label{app:gme}
This Appendix is organized as follows. First, we recap the framework of exact Gaussian master equations (GMEs) and apply it to describe the exact dynamics of the gQBM model. Then, we calculate the Keldysh components of the dressed Green's function, which is the core problem of the GME formalism. 
\subsection{Gaussian master equation formalism}
Let us consider a quantum system ($S$) coupled to an environment ($E$). The total Hamiltonian is $\H{} = \H{0} + \V = \H{S} + \H{E} + \V$, where the coupling is of the general bilinear form $\V = \sum_i A_i \otimes B_i$. Here, $\{A_i\}_i$ and $\{B_i\}_i$ denote system and environment operators, respectively. Adopting the Keldysh formalism, the system state $\rho_{S}(t) = \tr_E [\U(t, 0) \rho_{SE}(0) \U^\dagger(t, 0)]$ is expressed via the contour $\gamma(t)$, consisting of forward ($\gamma_-$) and backward ($\gamma_+$) branches ordered from $0$ to $t$ and $t$ to $0$, respectively [Fig.~\ref{smfig:1}]. Using the contour ordering operator $\tord$, which orders operators with earlier arguments in the contour to the right, the (interaction-picture) dynamics can be written in the compact form 
\begin{equation}
\label{s,-eq:state-keldysh-ord}
    \rho_S^{(I)}(t) = \tr_E\tord \left\{\exp\left[ -i\int_{\gamma(t)} \d z\,\V(z) \right] \rho_{SE}^{\vphantom{\dagger}}(0)\right\}~,
\end{equation}
where $\V(z)$ denotes the interaction-picture potential (see also \cite{cavina2025unifying}). 

Under the Gaussian hypothesis, $S$ and $E$ consist of non-interacting particles with quadratic Hamiltonians and bilinear coupling. For a separable initial state $\rho_{SE}(0) = \rho_S(0) \otimes \omega_E$, the dynamics is governed by the environment contour-valued Green's function (GF), i.e.
\begin{equation}
\label{sm-eq:bare-GF}
    \C_{ij}(z, w) := \tr\tord_\zeta\bigl[B_i^\dagger(z) B_j(w) \omega_E\bigr]~,
\end{equation}
where $\tord_\zeta$ accounts for quantum statistics ($\zeta = +1$ for bosons, $\zeta = -1$ for fermions). Correlated initial states can be treated similarly~\cite{d2025exact}. It has recently been shown~\cite{d2025exact} that the exact dynamics follows a Gaussian master equation (GME) (here written in interaction picture):
\begin{equation}
    \label{sm-eq:gme-general}
    \frac{\d \rho_S^{(I)}(t)}{\d t} = \sum_{i, j}\int_{0}^t \d \tau \, \G_{ij}^>(t, \tau) \bigl[{A_j(\tau) \rho_S^{(I)}(t)}, A_i^\dagger(t)\bigr] + \hc~,
\end{equation}
of which Eq.~\eqref{eq:gme-dissipator} of the main text constitutes the Schr\"odinger picture analogue. Here, $\G^>(t, \tau)$ is the greater Keldysh component of the contour-valued dressed Green's function (GF) $\G(t,\tau)$, i.e. $\G(\tau_1, \tau_2)$ with $\tau_1 \in \gamma_+(t)$ and $\tau_2 \in \gamma_-(t)$. The contour-valued GF satisfies the following Dyson equation:
\begin{equation}
\label{sm-eq:bare-dyson}
     \G(t, \tau) = \C(t, \tau) - \iint_{\gamma(t)} \d u \,\d v\, \G(t, u)\Sigma(u,v ) \C(v, \tau)~.
\end{equation}
In analogy with many-body theory~\cite{FetterWalecka, GiulianiVignale}, the self-energy $\Sigma$ encodes the memory of repeated system–environment interactions, accounting for vertices to all orders. At the lowest non-trivial order ($\G \to \C$), Eq.~\eqref{sm-eq:gme-general} reduces to the standard Redfield equation~\cite{d2025exact, breuer2007book, deVega2017review}. 
For the case of the gQBM model, the Hamiltonian is given by Eqs.~\eqref{eq:system-hamiltonian}, \eqref{eq:environment-hamiltonian}, \eqref{eq:interaction-hamiltonian}, in turns the system operators coincide with the quadratures $x,p$ and the GME in Schr\"odinger picture~\eqref{sm-eq:gme-general} can be cast in the form of Eq.~\eqref{eq:gme} given in the main text, i.e.
\begin{equation}
\label{sm-eq:gme}
    \frac{\d  \rho_S(t)}{ \d t} =-i\bigl[{\H{}}_S,  \rho_S(t)\bigr]+ \sum_{i,j} \int_0^t\d \tau\,\G^>_{ij}(t, \tau)\,\bigl[ r_j(\tau-t)  \rho_S(t), r_i\bigr] + \hc~,
\end{equation}
where 
\begin{align}
   r_1(t) &= x(t) := e^{i {\H{}}_S t} \, x\,e^{-i {\H{}}_S t} = {x} \cos(\omega_0 t) +  \frac{p}{m\omega_0} \sin(\omega_0 t)~,\\
    r_2(t) &= p(t) := e^{i {\H{}}_S t} \, p\,e^{-i {\H{}}_S t} = p \cos(\omega_0 t)-m\omega_0x \sin(\omega_0 t) ~,
\end{align}
%
and the self-energy $\Sigma$ appearing in Eq.~\eqref{sm-eq:bare-dyson} is given by
\begin{equation}
\label{sm-eq:self-energy}
     \Sigma_{i j}(z, w) : =
    \begin{cases}
        -\theta_{z \prec w} \bigl[ r_i(z),  r_j(w)\bigr] & {\rm if} \quad z \in \gamma_-(t)~,\\
        +\theta_{z \succ w} \bigl[ r_i(z),  r_j(w)\bigr] & {\rm if} \quad z \in \gamma_+(t)~.
    \end{cases}
\end{equation}
The dressed GF $\mathcal{G}$ can be obtained by solving Eq. \eqref{sm-eq:bare-dyson} once the correlation function $\mathcal{C}$, determined by the environmental operators in the interaction Hamiltonian, is specified. 
For the gQBM model, we have
\begin{align} \notag
    B_1(t) &:= \frac{1+\alpha}{2} X(t)= \frac{1+\alpha}{2} e^{i \H{E}t } X e^{-i \H{E}t} ~, \\
    B_2(t)& :=\frac{1-\alpha}{2} P(t)= \frac{1-\alpha}{2} e^{i \H{E}t } P e^{-i \H{E}t} ~, \label{eq:operators}
\end{align}
\begin{figure}
    \def\svgwidth{0.5\linewidth}
\begingroup%
  \makeatletter%
  \providecommand\color[2][]{%
    \errmessage{(Inkscape) Color is used for the text in Inkscape, but the package 'color.sty' is not loaded}%
    \renewcommand\color[2][]{}%
  }%
  \providecommand\transparent[1]{%
    \errmessage{(Inkscape) Transparency is used (non-zero) for the text in Inkscape, but the package 'transparent.sty' is not loaded}%
    \renewcommand\transparent[1]{}%
  }%
  \providecommand\rotatebox[2]{#2}%
  \newcommand*\fsize{\dimexpr\f@size pt\relax}%
  \newcommand*\lineheight[1]{\fontsize{\fsize}{#1\fsize}\selectfont}%
  \ifx\svgwidth\undefined%
    \setlength{\unitlength}{198.70299091bp}%
    \ifx\svgscale\undefined%
      \relax%
    \else%
      \setlength{\unitlength}{\unitlength * \real{\svgscale}}%
    \fi%
  \else%
    \setlength{\unitlength}{\svgwidth}%
  \fi%
  \global\let\svgwidth\undefined%
  \global\let\svgscale\undefined%
  \makeatother%
  \begin{picture}(1,0.3908932)%
    \lineheight{1}%
    \setlength\tabcolsep{0pt}%
    \put(0,0){\includegraphics[width=\unitlength,page=1]{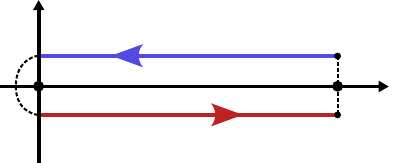}}%
    \put(0.36637574,0.28587962){\makebox(0,0)[lt]{\lineheight{1.10000002}\smash{\begin{tabular}[t]{l}$\gamma_+(t)$\end{tabular}}}}%
    \put(0.36637574,0.0697289){\makebox(0,0)[lt]{\lineheight{1.10000002}\smash{\begin{tabular}[t]{l}$\gamma_-(t)$\end{tabular}}}}%
    \put(0.82710196,0.20689213){\makebox(0,0)[lt]{\lineheight{1.10000002}\smash{\begin{tabular}[t]{l}$t$\end{tabular}}}}%
    \put(0.79374804,0.07204543){\makebox(0,0)[lt]{\lineheight{1.10000002}\smash{\begin{tabular}[t]{l}$t_-$\end{tabular}}}}%
    \put(0.79374804,0.28308638){\makebox(0,0)[lt]{\lineheight{1.10000002}\smash{\begin{tabular}[t]{l}$t_+$\end{tabular}}}}%
    \put(0.11068442,0.20092514){\makebox(0,0)[lt]{\lineheight{1.10000002}\smash{\begin{tabular}[t]{l}$0$\end{tabular}}}}%
    \put(0.11551738,0.366858){\makebox(0,0)[lt]{\lineheight{1.10000002}\smash{\begin{tabular}[t]{l}$\Im z$\end{tabular}}}}%
    \put(0.898463,0.21542423){\makebox(0,0)[lt]{\lineheight{1.10000002}\smash{\begin{tabular}[t]{l}$\Re z$\end{tabular}}}}%
  \end{picture}%
\endgroup%

    \caption{Graphic depiction of the Schwinger-Keldysh contour in the complex time plane.}
    \label{smfig:1}
\end{figure}
and the bare GF in Eq.~\eqref{sm-eq:bare-GF} is thus given by
\begin{equation}
    \C(z, w) := \frac{1}{4}
    \begin{pmatrix}
        (1+\alpha)^2\tr\bigl[\tord\bigl\{ X(z) X(w)\omega_E\bigr\}\bigr] & (1-\alpha^2)\tr\bigl[\tord\bigl\{ X(z) P(w)\omega_E\bigr\}\bigr] \\[5pt]
        (1-\alpha^2)\tr\bigl[\tord\bigl\{ P(z) X(w)\omega_E\bigr\}\bigr] & (1-\alpha)^2\tr\bigl[\tord\bigl\{ P(z) P(w)\omega_E\bigr\}\bigr] 
    \end{pmatrix}~.
\end{equation}
We assume the environment to be prepared in a Gibbs state, i.e.
\begin{equation}
    \omega_E = \frac{e^{-\beta \H{E}}}{\tr \bigl[e^{-\beta \H{E}}\bigr]}~.
\end{equation}
\subsection{Calculation of the bare and dressed Green functions 
}

The environment quadratures, in terms of creation and annihilation operators $b_n^\dagger$, $b_n^{\vphantom{\dagger}}$, are given by
\begin{align}
    x_n(t) & = \frac{1}{\sqrt{2M_n\omega_n}} \Bigl(\b{n} e^{-i\omega_n t} +\bd{n} e^{i\omega_n t }\Bigr)\\
    p_n(t) &= -i \sqrt{\frac{M_n \omega_n}{2}} \Bigl(\b{n} e^{-i\omega_n t} - \bd{n} e^{i\omega_n t }\Bigr)~,
\end{align}
where $n$ labels the $n$-th environment mode of frequency $\omega_n$, and $M_n$ is the mass of each mode. The (interaction-picture) environment operators assume the form
\begin{align}
    {B}_1(t)& = \frac{1+\alpha}{2} X(t)= \frac{1+\alpha}{2}\sum_n f_n {x}_n(t) = \frac{1+\alpha}{2}\sum_n \sqrt{\frac{f_n^2}{2M_n\omega_n}} \Bigl(\b{n} e^{-i\omega_n t} +\bd{n} e^{i\omega_n t }\Bigr)~,\\
    {B}_2(t)& = \frac{1-\alpha}{2}P(t)= \frac{1-\alpha}{2}\sum_n g_n {p}_n(t) =-i\frac{1-\alpha}{2}\sum_n  \sqrt{\frac{g_n^2M_n \omega_n}{2}} \Bigl(\b{n} e^{-i\omega_n t} - \bd{n} e^{i\omega_n t }\Bigr)~.
\end{align}
Defining the physical-time GF
\begin{equation}
\label{eq:chi}
    \chi_{ij}(t) := \tr\bigl[{B}_i^\dagger(t) {B}_j(0) \omega_E\bigr]~,
\end{equation}
which is immediately seen to be homogeneous in $t$, we can obtain the Keldysh components of the contour GF as
\begin{align}
\label{eq:C>}
    \C^>(t) & = \chi(t)~,\\
\label{eq:C<}
    \C^<(t) & = \chi^T(-t)~,\\
\label{eq:Ctord}    
    \C^\tord(t) & = \theta(t)\chi(t) +\theta(-t)\chi^T(-t)~,\\
\label{eq:Cdrot}    
    \C^\drot(t) & =  \theta(-t)\chi(t) +\theta(t)\chi^T(-t)~.
\end{align}
We observe that, by definition~\eqref{eq:chi}, $\chi^T(-t) = \chi^*(t)$. For $\chi$, we have
\begin{align}
    \chi_{xx}(t) &:= \frac{(1+\alpha)^2}{4}\tr\bigl[X(t) X(0)\omega_E\bigr] = \frac{(1+\alpha)^2}{4}\int_0^{+\infty} d\omega J_{XX}(\omega)\Bigl[\coth\left(\frac{\beta \omega }{2}\right) \cos(\omega t) - i \sin(\omega t)\Bigr]~, \label{eq:chixx} \\[5pt]
    \chi_{xp}(t) &:= \frac{1-\alpha^2}{4} \tr\bigl[X(t) P(0)\omega_E\bigr] = \frac{1-\alpha^2}{4}\int_0^{+\infty} d\omega J_{XP}(\omega)\Bigl[\coth\left(\frac{\beta \omega }{2}\right) \sin(\omega t) + i \cos(\omega t)\Bigr]~, \label{eq:chixp} \\[5pt]
    \chi_{px}(t) &:= \frac{1-\alpha^2}{4}\tr\bigl[P(t) X(0)\omega_E\bigr]= -\frac{1-\alpha^2}{4}\int_0^{+\infty} d\omega J_{XP}(\omega)\Bigl[\coth\left(\frac{\beta \omega }{2}\right) \sin(\omega t) + i \cos(\omega t)\Bigr]~, \label{eq:chipx} \\[5pt]
    \chi_{pp}(t) &:= \frac{(1-\alpha)^2}{4} \tr\bigl[P(t) P(0)\omega_E\bigr] = \frac{(1-\alpha)^2}{4} \int_0^{+\infty} d\omega J_{PP}(\omega)\Bigl[\coth\left(\frac{\beta \omega }{2}\right) \cos(\omega t) - i \sin(\omega t)\Bigr]~. \label{eq:chipp}
\end{align}
where we defined the densities of states (DOSs)
\begin{align}
    J_{XX}(\omega) & :=\sum_n \frac{ f_n^2}{2M_n\omega_n} \, \delta(\omega - \omega_n) ~,\\
    J_{XP}(\omega) & :=\sum_n \frac{ f_n g_n}{2} \, \delta(\omega - \omega_n) ~,\\
    J_{PP}(\omega) & :=\sum_n \frac{g_n^2 M_n \omega_n }{2} \, \delta(\omega - \omega_n) ~,
\end{align}
In the following we will assume that the coupling constant $f_n$ and $g_n$ are chosen in such a way that $f_n = M_n \omega_n g_n$ $\forall n$, so that $J_{XX} \equiv J_{XP} \equiv J_{PP} := J$. In this way, the relative weight of the $x \otimes X$ and $p\otimes P$ coupling channels on the dynamics is uniquely quantified by the dimensionless parameter $\alpha$. Thus, we have compactly
\begin{align}
\label{sm-eq:re-chi}
    \Re \chi(t) & = \frac{1}{4}\int_{0}^{\infty} \d \omega \,J(\omega)
    \coth\left(\frac{\beta \omega}{2}\right) 
    \begin{pmatrix}
        (1+\alpha)^2\cos(\omega t) & (1-\alpha^2) \sin(\omega t)\\[5pt]
         -(1-\alpha^2) \sin(\omega t) & (1-\alpha)^2\cos(\omega t)
    \end{pmatrix}~,\\[8pt]
\label{sm-eq:im-chi}
    \Im \chi (t) & = -\frac{1}{4}\int_{0}^{\infty} \d \omega \,J(\omega)\begin{pmatrix}
        (1+\alpha)^2\sin(\omega t) & -(1-\alpha^2) \cos(\omega t)\\[5pt]
         (1-\alpha^2) \cos(\omega t) & (1-\alpha)^2\sin(\omega t)
    \end{pmatrix}~.
\end{align}
The real and imaginary parts of $\chi(t)$ computed above are directly connected to specific combinations of the bare Green's function components. 
If we define the retarded (R), advanced (A) and Keldysh (K) components of $\mathcal{C}(t)$ as the ones satisfying
\begin{align}
    \C^R - \C^A = - i(\C^> - \C^<) ~,\qquad
    \C^K = \C^> +\C^< ~.
\end{align}
with $\C^R(t, \tau) \neq 0$ only for $t \geq \tau$ and $\C^A(t, \tau) \neq 0$ for $t < \tau$, it is easy to verify that
\begin{equation} \label{eq:C-to-chi}
    \C^K(s) = 2 \Re \chi(s)~, \quad \C^R(s) = 2\theta(s) \Im \chi(s)~, \quad \C^A(s) = -2\theta(s) \Im \chi(s)~.
\end{equation}
An analogous expression can be used for the dressed Green's function $\mathcal{G}$, for which we have
\begin{align}
    \G^K  := 2 \Re\G^>~,\qquad
    \G^R - \G^A  := 2 \Im\G^>~,
\end{align}
with $\G^R(t, \tau) \neq 0$ only for $t \geq \tau$ and $\G^A(t, \tau) \neq 0$ for $t < \tau$. Thus, we have
\begin{align}
    \G^>(t, \tau) = \bigl[\G^<(t, \tau)\bigr]^* = \frac{\G^K(t, \tau)+i\G^R(t, \tau)-i\G^A(t, \tau)}{2} = \frac{\G^K(t, \tau)+i\G^R(t, \tau)}{2}~,
\end{align}
where the last equality holds only for $t \geq \tau$.
Let us turn our attention to the (retarded) self-energy $\Sigma^R = -i\Sigma^\tord$. It is immediately proven to be homogeneous and can be expressed compactly as
\begin{equation}
    \Sigma^R_{ij}(t) =\theta(-t) \sigma_{ij} (t) := \theta(-t) i\bigl[ r_i(t),  r_j(0)\bigr]~,
\end{equation}
and we straightforwardly find
\begin{align}
    \sigma_{xx}(t) & := i\Bigl[{x} \cos(\omega_0 t) +  \frac{p}{m\omega_0} \sin(\omega_0 t),  x\Bigr] = +\frac{\sin(\omega_0 t)}{m\omega_0}~,\\
    \sigma_{xp}(t) & := i\Bigl[{x} \cos(\omega_0 t) +  \frac{p}{m\omega_0} \sin(\omega_0 t),  p\Bigr] = -\cos(\omega_0 t)~,\\
    \sigma_{px}(t) & := i\Bigl[{p} \cos(\omega_0 t) +  m\omega_0 x \sin(\omega_0 t),  x\Bigr] = +\cos(\omega_0 t)~,\\
    \sigma_{pp}(t) & :=  i\Bigl[{p} \cos(\omega_0 t) +  m\omega_0 x \sin(\omega_0 t),  p\Bigr] = -\frac{\sin(\omega_0 t)}{m\omega_0}~.
\end{align}
Compactly,
\begin{equation}
\label{sm-eq:sigma_retarded}
    \Sigma^R(t) = \theta(-t)
    \begin{pmatrix}
        \frac{\sin(\omega_0 t)}{m\omega_0} & -\cos(\omega_0 t)\\[5pt]
        \cos(\omega_0 t) & -\frac{\sin(\omega_0 t)}{m\omega_0}
    \end{pmatrix}~.
\end{equation}
We note that, technically, this “retarded" self-energy is not defined on the usual characteristic dominion of the retarded components of contour-valued functions~\cite{breuer2007book, kamenev2011book}, which would be $t\geq 0$, and for the advanced component we have $\Sigma^A = \Sigma^R$. Furthermore, introducing an extra $-i$ factor---consistently with the definition of $\G^R$---ensures that both $\G^R$ and $\G^K$ are real-valued functions, facilitating analytical and numerical calculations. The great advantage of using $\G^{R}$, $\G^{A}$, $\G^{K}$ instead of the lesser, greater, time-ordered and anti-time-ordered components, together with this definition of the self-energy, is the simplification that occurs in the Dyson equation. 
Indeed, the following physical-time, real-valued Dyson forms hold for the dressed GFs can be derived from the contour equation~\eqref{sm-eq:bare-dyson}:
\begin{align}
\label{eq:dyson-retarded}
    \G^R & = \C^R + \G^R \circ \Sigma^R \circ\C^R~,\\
\label{eq:dyson-advanced}
    \G^A & = \C^A + \G^A \circ \Sigma^A \circ\C^A~,\\
\label{eq:dyson-keldysh}
    \G^K & = \bigl[\1 +\G^R\circ \Sigma^R \bigr]\circ\C^K +\G^K \circ \Sigma^R \circ \C^A~,
\end{align}
where we employed the shorthand notation
\begin{equation}
    f \circ g \circ h := \iint_{[0,t]} \d u\,\d v\, f(t, u) g(u,v) h(v, \tau)~.
\end{equation}
This means that to obtain $\G^{>}$, that appear as the memory kernel of the GME, it will be sufficient to solve Eqs.~\eqref{eq:dyson-retarded},~\eqref{eq:dyson-advanced},~\eqref{eq:dyson-keldysh}, that are defined in physical time rather than through integrals on the Keldysh contour (like Eq.~\eqref{sm-eq:bare-dyson}).
From a practical point of view we will obtain the retarded GF numerically, as the fixed point of the linear map defined by Eq.~\eqref{eq:dyson-retarded}, as detailed in App.~\ref{app:parameters}.
The Keldysh component can be obtained in the same way from Eq.~\eqref{eq:dyson-keldysh}, after the values of $\G^R$ are known.
\section{Quasi-GKSL form of the Gaussian master equation}
\setcounter{figure}{0}
\label{app:gksl}
In this appendix, we start by providing a general derivation of the quasi-GKSL form of the GME dissipator. Then, we apply this general construction to the gQBM model, and numerically prove the non-negativity of the trace of the Kossakowski matrix for the gQBM.
%
%
To write the GME in the quasi-GKSL form [Eq.~\eqref{eq:quasi-gksl} of the main text] we start by expanding the interaction-picture dissipator of the GME [Eq.~\eqref{eq:gme-dissipator} of the main text] in terms of the static quadratures, i.e.
\begin{equation}
    A_i(t) = \sum_{j}\AA_{ij}(t) r_j~.
\end{equation}
Moreover, we write the system Hamiltonian as a quadratic form of the coordinates, i.e.
\begin{equation}
    \label{sm-eq:hmatrix}
    \H{S}  := \frac{1}{2}\sum_{i,j} h_{ij} r_i r_j~.
\end{equation}
Thus, we have
\begin{equation}
    \begin{split}
        \frac{\d  \rho_S}{ \d t} & =  - \frac{i}{2} h_{ij}\bigl[r_i r_j, \rho_S\bigr] + \M_{ij}(t)\bigl[ r_j  \rho, r_i\bigr] +\M_{ij}^*(t)\bigl[ r_i, \rho_S r_j\bigr]~,
    \end{split}
\end{equation}
where the sum on repeated indices is implied, and we defined
\begin{equation}
\label{sm-eq:M-matrix}
    \M(t):= \int_0^t\d \tau\, \AA^\dagger(0)\G^>(t, \tau)\AA(\tau-t)~,
\end{equation}
as in Eq.~\eqref{eq:M-matrix} of the main text.
The following identities hold:
\begin{align}
    \bigl[ r_j  \rho_S, r_i\bigr] & = r_j \rho_S r_i - r_i r_j \rho_S = r_j \rho_S r_i - \frac{1}{2}\bigl\{r_i r_j, \rho_S\bigr\} - \frac{1}{2}\bigl[r_i r_j, \rho_S\bigr]~, \\
    \bigl[ r_i, \rho_S r_j \bigr] & = r_i \rho_S r_j - \rho_S r_j r_i = r_i \rho_S r_j - \frac{1}{2}\bigl\{r_j r_i, \rho_S\bigr\} + \frac{1}{2}\bigl[r_j r_i, \rho_S\bigr]~.
\end{align}
The general quasi-GKSL form, with the static quadratures as Lindblad operators, is
\begin{equation}
\label{eq:quasi-gksl}
    \frac{\d  \rho_S}{\d t} = -i\bigl[\Lambda(t),  \rho_S\bigr] + \sum_{i,j} \gamma_{ij}(t) \D_{ij}\rho_S~,
\end{equation}
with
\begin{equation}
    \D_{ij}\,\bullet := r_i \bullet r_j - \frac{1}{2}\bigl\{r_j r_i, \bullet\bigr\}~.
\end{equation}
With some relabeling, we have
\begin{equation}
\label{eq:expansion}
    \begin{split}
         \dot \rho_S  
         & =  - \frac{i}{2} h_{ij}\bigl[r_i r_j, \rho_S\bigr] +\M_{ij}(t) \Bigl(\D_{ji} \rho_S - \frac{1}{2}\bigl[r_i r_j, \rho_S\bigr]\Bigr) + \M_{ij}^*(t)\Bigl(\D_{ij} \rho_S + \frac{1}{2}\bigl[r_j r_i, \rho_S\bigr]\Bigr) \\
         & =   - \frac{i}{2} h_{ij}\bigl[r_i r_j, \rho_S\bigr] +\M_{ji}(t) \Bigl(\D_{ij} \rho_S - \frac{1}{2}\bigl[r_j r_i, \rho_S\bigr]\Bigr) + \M_{ij}^*(t)\Bigl(\D_{ij} \rho_S + \frac{1}{2}\bigl[r_j r_i, \rho_S\bigr]\Bigr)\\
         & =  \bigl[\M_{ji}(t)+ \M_{ij}^*(t)\bigr]  \D_{ij} \rho_S -\frac{i}{2} \,\Bigl(h_{ij}+i\bigl[\M_{ij}^*(t)-\M_{ji}(t) \bigr]\Bigr)\bigl[r_j r_i, \rho_S\bigr]~.
    \end{split}   
\end{equation}
Thus, comparing Eqs.~\eqref{eq:expansion} and~\eqref{eq:quasi-gksl}, we find the Kossakowski matrix to be
\begin{align}
\label{sm-eq:m+mt}
    \gamma(t) & = \M^*(t)+\M^T(t)~.
\end{align}
The Lamb-shifted Hamiltonian can be parametrized as a quadratic form in the quadratures, i.e.
\begin{equation}
    \Lambda(t) := \frac{1}{2}\bm r^T \cdot \bigl[h + \delta(t)\bigr]\cdot  \bm r~,
\end{equation}
with
\begin{align}
    \delta(t)& := i \bigl[\M^*(t)- \M^T(t)\bigr]~.
\end{align}
The manipulations above holds in general for the dynamics of any bosonic Gaussian system, expressing the GME in the quasi-GKSL form.
For the case of the gQBM model, fixing $m = 1 =\omega_0$ for simplicity, it is easy to verify that
\begin{equation}
    \AA(t) := 
    \begin{pmatrix}
        \cos(t) &\sin(t) \\[5pt]
          -\sin(t)& \cos( t) 
    \end{pmatrix}~.
\end{equation}
Thus, the matrix $\M(t)$ defined in Eq.~\eqref{sm-eq:M-matrix} becomes
\begin{equation}
    \M(t) = \int_{0}^t \d \tau 
    \begin{pmatrix}
        \G^>_{xx}(t, \tau) \cos(\tau-t) - \G^>_{xp}(t, \tau) \sin(\tau -t) & \G^>_{xx}(t, \tau) \sin(\tau-t) + \G^>_{xp}(t, \tau) \cos(\tau -t)\\[5pt]
        \G^>_{px}(t, \tau) \cos(\tau-t) - \G^>_{pp}(t, \tau) \sin(\tau -t) & \G^>_{px}(t, \tau) \sin(\tau-t) + \G^>_{pp}(t, \tau) \cos(\tau -t)
    \end{pmatrix}
\end{equation}
The Kossakowski matrix elements, according to Eq.~\eqref{sm-eq:m+mt}, are thus given by
\begin{align}
\label{sm-eq:gammaxx}
    \gamma_{xx}(t) & = \int_0^t \d \tau\Bigl[\G^K_{xx}(t,\tau)\cos(\tau-t) - \G^K_{xp}(t,\tau)\sin(\tau-t)\Bigr] ~,\\
\label{sm-eq:gammapp}
    \gamma_{pp}(t) & = \int_0^t \d \tau\Bigl[  \G^K_{pp}(t, \tau)\cos(\tau-t) +\G^K_{px}(t, \tau)\sin(\tau-t)\Bigr] ~,\\
\label{sm-eq:gammaxp}
    \gamma_{xp}(t) & =\int_{0}^t\d\tau \Bigl[\bigl[\G_{xp}^{>*}(t,\tau) + \G_{px}^{>}(t,\tau)\bigr]\cos(\tau-t) + \bigl[\G_{xx}^{>*}(t,\tau) - \G_{pp}^>(t,\tau)\bigr]\sin(\tau-t) \Bigr]~,
\end{align}
and $\gamma_{px}^{\vphantom{*}} = \gamma_{xp}^*$, with real and imaginary parts given by
\begin{align}
\label{sm-eq:regammaxp}
    \Re \gamma_{xp}(t) & =\frac{1}{2}\int_{0}^t\d\tau \Bigl[\bigl[\G_{xp}^K(t,\tau)+\G_{px}^{K}(t,\tau)\bigr]\cos(\tau-t) + \bigl[\G_{xx}^{K}(t,\tau) - \G_{pp}^K(t,\tau)\bigr]\sin(\tau-t) \Bigr]~,\\
\label{sm-eq:imgammaxp}
    \Im \gamma_{xp}(t) & =\frac{1}{2}\int_{0}^t\d\tau \Bigl[\bigl[\G_{px}^R(t,\tau)-\G_{xp}^{R}(t,\tau)  \bigr]\cos(\tau-t) - \bigl[\G_{xx}^{R}(t,\tau) +\G_{pp}^R(t,\tau)\bigr]\sin(\tau-t) \Bigr]~.
\end{align}
\section{Parameters of the gQBM and numerical analysis}
\label{app:parameters}
With our choice of coupling parameters, the influence of the environment on the system is completely determined by a single function, the spectral density $J(\omega)$ (see Eqs.~\eqref{sm-eq:im-chi}, \eqref{sm-eq:im-chi}).
Throughout this work, we employ a sub-ohmic spectral density with an exponential cutoff
\begin{equation}
\label{sm-eq:dos}
    J(\omega) = \eta \omega^s e^{-\omega/\omega_c}~,
\end{equation}
with $\eta = 0.1$, $s = 0.8$ and $\omega_c = \omega_0 = 1$. Unless otherwise indicated, the inverse temperature of the environment, entering through the Keldysh Green function as in~\eqref{sm-eq:re-chi}, is set to $\beta = 1/2\omega_0 = 1/2$. 
As discussed in Appendix~\ref{app:gksl}, the first step in the calculation of the Kossakowski matrix is the numerical evaluation of $\mathcal{G}^R$, $\mathcal{G}^K$ through Eqs.~\eqref{eq:dyson-keldysh}, \eqref{eq:dyson-retarded}, given the components of the self-energy $\Sigma$ and of the correlation function $\C$ (see Eqs.~\eqref{eq:C-to-chi},~\eqref{sm-eq:sigma_retarded}). This is achieved by finding the fixed point of the following dynamical maps:
\begin{equation}
    \begin{cases}
        \G^R_{0} = \C^R~,\\
        \G^R_n = \C^R + \G^R_{n-1} \circ \Sigma^R \circ\C^R~,\\
    \end{cases}
\end{equation}
\begin{equation}
    \begin{cases}
        \G^K_{0} = \C^K +\G^R\circ \Sigma^R \circ\C^K~,\\
        \G^K_{n} = \G^K_{0} + \G_{n-1}^K \circ \Sigma^R \circ \C^A~.
    \end{cases}
\end{equation}
The maps are applied iteratively $N$ times until we achieve convergence in Frobenius norm $\|\bullet\|_2$ within a certain relative tolerance $\epsilon$, i.e. 
\begin{equation}
    \Bigl\|\G^{R/K}_{N}- \G^{R/K}_{N-1}\Bigr\|_2 := \left(\sum_{i,j} \int_{0}^t \d u \int_{0}^{u} \d v\, \Bigl|\bigl[\G^{R/K}_{N}(u, v)\bigr]_{ij}- \bigl[\G^{R/K}_{N-1}(u, v)\bigr]_{ij}\Bigr|^2\right)^{1/2}  < \epsilon \bigl\|\G^{R/K}_{N}\bigr\|_2~.
\end{equation}
In this work, we have fixed $\epsilon = 10^{-7}$.

After $\mathcal{G}^K$ and $\mathcal{G}^R$ are known, the Kossakowski matrix is obtained for the gQBM by substitution in Eqs.~\eqref{sm-eq:gammaxx}, \eqref{sm-eq:gammapp}, \eqref{sm-eq:gammaxp}.
 In general, according to the criterion given in Eq. \eqref{eq:cp-criterion}, we are interested in checking the positive-semidefiniteness of the Kossakowski matrix, that can be done by computing its eigenvalues, that are
\begin{equation}
    \gamma_{\pm}(t) = \frac{1}{2} \biggl[\gamma_{xx}(t) + \gamma_{pp}(t) \pm \sqrt{\bigl[\gamma_{xx}(t)-\gamma_{pp}(t)\bigr]^2 + 4\bigl|\gamma_{xp}(t)\bigr|^2}\,\biggr]~.
\end{equation}
Each of these eigenvalues is a canonical rate associated with a particular dissipator, as discussed in the main text and computed analytically in the low-temperature regime.
However, for the purpose of assessing the positive-semidefiniteness of $\gamma(t)$, it is more immediate to use the conditions
\begin{equation}
    \begin{cases}
        \gamma_{xx}(t) + \gamma_{pp}(t) \geq 0~, \\
        \gamma_{xx}(t)\gamma_{pp}(t) \geq |\gamma_{xp}(t)|^2 ~,
    \end{cases}
\end{equation}
indeed, the first condition ($\tr \gamma(t) \geq 0$) is easily accessible numerically, and as shown in Fig.~\ref{smfig:2}, it is satisfied for every value of the control parameter $\alpha$ explored in this work, so the sufficient and necessary condition for positive-semidefiniteness of $\gamma(t)$ is effectively given by $\det \gamma(t) \geq 0$.
\begin{figure}
    \centering
    \begin{overpic}[width = 0.9\linewidth]{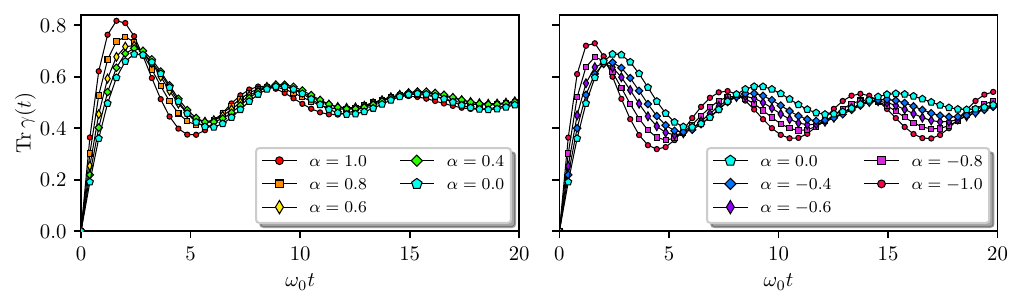}
        \put(1,29){(a)}
        \put(51.5,29){(b)}
    \end{overpic}
    \caption{Trace of the Kossakowski matrix $\gamma(t)$ as a function of time for different values of the control parameter $\alpha$. (a) $\alpha \in [0, 1]$. (b) $\alpha \in [-1, 0]$. We observe that $\tr \gamma(t) \geq 0$ $\forall t$ for every value of the control parameter $\alpha$. Other numerical parameters are fixed as in App.~\ref{app:parameters}.}
    \label{smfig:2}
\end{figure}
\section{Non-Markovianity from symmetry-breaking terms}
\label{app:crt}

To understand how CRTs control memory effects in the gQBM, it is sufficient to perform a weak-coupling analysis. Starting from Eq.~\eqref{eq:interaction-hamiltonian2} of the main text, we want to rewrite the gQBM dissipator in terms of creation and annihilation operators. We have system operators $A_1 = \a{}$, $A_2 = \ad{}$ and environment operators $B^\dagger + \alpha B$, $B + \alpha B^\dagger$. The physical-time correlator is
\begin{equation}
    \begin{split}
    \chi(s) &= 
    \begin{pmatrix}
       \vev{B(s) B^\dagger(0)} + \alpha^2 \vev{B^\dagger(s) B(0)} &  \alpha \Bigl[ \vev{B^\dagger(s) B(0)} + \vev{B(s) B^\dagger(0)}\Bigr] \\[5pt]
         \alpha \Bigl[ \vev{B^\dagger(s) B(0)} + \vev{B(s) B^\dagger(0)}\Bigr] &  \vev{B^\dagger(s) B(0)} + \alpha^2 \vev{B(s) B^\dagger(0)}
    \end{pmatrix}\\[5pt]
    & 
    :=
    \begin{pmatrix}
        \chi_+(s) + \alpha^2 \chi_-(s)  & \alpha \bigl[\chi_-(s) + \chi_+(s)\bigr] \\[5pt]
        \alpha \bigl[\chi_-(s) + \chi_+(s)\bigr]& \chi_-(s) + \alpha^2 \chi_+(s)
    \end{pmatrix}~,
    \end{split}
\end{equation}
with
\begin{equation}
    \begin{split}
        \chi_-(s) 
        & := \vev{B^\dagger(s) B(0)} = \sum_{n,m} k_n^* k_m  e^{i\omega_n s} \vev{\ad{n} \a{m}} = \sum_n |k_n|^2 \delta(\omega-\omega_n)e^{i\omega_n s} \vev{\ad{n} \a{n}}\\
        & = \int_0^\infty \d \omega\,J(\omega) n(\omega) e^{i\omega s}~,
    \end{split}
\end{equation}
and
\begin{equation}
    \begin{split}
        \chi_+(s) 
        & := \vev{B(s) B^\dagger(0)} = \sum_{n,m} k_n k_m^*  e^{-i\omega_n s} \vev{\a{n} \ad{m}} = \sum_n |k_n|^2 \delta(\omega-\omega_n)e^{-i\omega_n s} \vev{\a{n} \ad{n}}\\
        & = \int_0^\infty \d \omega\,J(\omega) \bigl[n(\omega) +1 \bigr]e^{-i\omega s}~,
    \end{split}
\end{equation}
where $k_n := \sqrt{f_n g_n/2}$, and $n(\omega) := \bigl[e^{\beta \omega}-1\bigr]^{-1}$ is the Bose-Einstein distribution.
Let's calculate the Kossakowski matrix in the weak-coupling limit. We have
\begin{equation}
    \begin{pmatrix}
        \a{}(t) \\
        \ad{}(t)
    \end{pmatrix}
    =
    \AA(t) 
    \begin{pmatrix}
        \a{} \\
        \ad{}
    \end{pmatrix}
    = 
    \begin{pmatrix}
        e^{-i\omega_0 t} & 0 \\
        0 & e^{+i\omega_0t}
    \end{pmatrix}
    \begin{pmatrix}
        \a{} \\
        \ad{}
    \end{pmatrix}~,
\end{equation}
and $\G^> = \chi + \O(\eta^2)$, where $\eta$ is defined in Eq.~\eqref{sm-eq:dos}. Thus,
\begin{equation}
    \begin{split}
        \M(t) 
        & = \int_0^t \d \tau\, \AA^T(0) \chi(t-\tau) \AA(\tau-t) \\
        & =  \int_0^t \d \tau\,
        \begin{pmatrix}
             \bigl[\chi_+(t-\tau) + \alpha^2 \chi_-(t-\tau) \bigr] e^{-i\omega_0(\tau-t)}  & \alpha \bigl[\chi_-(t-\tau) + \chi_+(t-\tau)\bigr]e^{+i\omega_0(\tau-t)} \\[5pt]
       \alpha \bigl[\chi_-(t-\tau) + \chi_+(t-\tau)\bigr]e^{-i\omega_0(\tau-t)}  & \bigl[\chi_-(t-\tau) + \alpha^2 \chi_+(t-\tau) \bigr]e^{+i\omega_0(\tau-t)}
        \end{pmatrix}~,
    \end{split}
\end{equation}
up to corrections of $\O(\eta^2)$.
Since we have
\begin{align}
    \int_{0}^t \d \tau \,e^{\pm i (\omega \pm \omega_0) (\tau-t)} = 2 e^{\mp i(\omega \pm \omega_0)t/2} \frac{\sin[(\omega \pm\omega_0)t/2]}{\omega \pm \omega_0}~,
\end{align}
the $\M$ matrix is, element-by-element, 
\begin{align}
    \M_{11}(t) & = 2\int_{0}^\infty \d \omega\,J(\omega) \Bigl(\alpha^2 n(\omega) e^{+ i(\omega + \omega_0)t/2} \frac{\sin[(\omega +\omega_0)t/2]}{\omega + \omega_0} + \bigl[n(\omega) +1 \bigr] e^{- i(\omega - \omega_0)t/2} \frac{\sin[(\omega -\omega_0)t/2]}{\omega - \omega_0} \Bigr)~,\\
    \M_{12}(t) & = 2\alpha\int_{0}^\infty \d \omega\,J(\omega) \,\Bigl(n(\omega) e^{+ i(\omega - \omega_0)t/2} \frac{\sin[(\omega -\omega_0)t/2]}{\omega - \omega_0} + \bigl[n(\omega) +1 \bigr]e^{- i(\omega + \omega_0)t/2} \frac{\sin[(\omega +\omega_0)t/2]}{\omega + \omega_0} \Bigr)~,\\
    \M_{21}(t) & = 2\alpha\int_{0}^\infty \d \omega\,J(\omega) \,\Bigl(n(\omega) e^{+ i(\omega + \omega_0)t/2} \frac{\sin[(\omega +\omega_0)t/2]}{\omega + \omega_0}  + \bigl[n(\omega) +1 \bigr] e^{-i(\omega - \omega_0)t/2} \frac{\sin[(\omega -\omega_0)t/2]}{\omega - \omega_0} \Bigr)~,\\
    \M_{22}(t) & = 2\int_{0}^\infty \d \omega\,J(\omega) \Bigl(n(\omega) e^{+ i(\omega - \omega_0)t/2} \frac{\sin[(\omega -\omega_0)t/2]}{\omega - \omega_0}  + \alpha^2\bigl[n(\omega) +1 \bigr] e^{- i(\omega + \omega_0)t/2} \frac{\sin[(\omega +\omega_0)t/2]}{\omega + \omega_0} \Bigr)~.
\end{align}
For the Kossakowski matrix, we thus have
\begin{align}
    \gamma_{11}(t) & = 2\int_{0}^\infty \d \omega\,J(\omega) \left[ \bigl[n(\omega) +1 \bigr] \frac{\sin[(\omega -\omega_0)t]}{\omega - \omega_0} + \alpha^2 n(\omega)\frac{\sin[(\omega +\omega_0)t]}{\omega + \omega_0}\right]~,\\
    \gamma_{22}(t) & = 2\int_{0}^\infty \d \omega\,J(\omega) \left[ n(\omega)\frac{\sin[(\omega -\omega_0)t]}{\omega - \omega_0} + \alpha^2  \bigl[n(\omega) +1 \bigr] \frac{\sin[(\omega +\omega_0)t]}{\omega + \omega_0}\right]~,\\
    \gamma_{12}(t) & = 2\alpha\int_{0}^\infty \d \omega\,J(\omega) \coth \left(\frac{\beta \omega}{2}\right) \,\left[e^{-i(\omega - \omega_0)t/2}\frac{\sin[(\omega -\omega_0)t/2]}{\omega - \omega_0} + e^{+i(\omega + \omega_0)t/2}\frac{\sin[(\omega +\omega_0)t/2]}{\omega + \omega_0} \right]~,
\end{align}
and $\gamma_{21} = \gamma_{12}^*$, with
\begin{align}
    \Re \gamma_{12}(t) & = \alpha \int_{0}^\infty \d \omega\,J(\omega) \coth \left(\frac{\beta \omega}{2}\right) \left[\frac{\sin[(\omega -\omega_0)t]}{\omega- \omega_0} +\frac{\sin[(\omega +\omega_0)t]}{\omega +\omega_0}\right]~, \\
    \Im \gamma_{12}(t) & = -2\alpha \int_{0}^\infty \d \omega\,J(\omega) \coth \left(\frac{\beta \omega}{2}\right) \left[\frac{\sin^2[(\omega -\omega_0)t/2]}{\omega- \omega_0}-  \frac{\sin^2[(\omega +\omega_0)t/2]}{\omega +\omega_0}\right]~.
\end{align}
To get some insight on the possibility of having $\gamma \ngeq 0$ for $t \to \infty$ (the case of eternal non-Markovianity) we may take the distributional limits
\begin{align}
    \lim_{t \to \infty} \int_0^{\infty} \d \omega \,\frac{\sin[(\omega -\omega_0)t]}{\omega- \omega_0} \phi(\omega) &= \pi \phi(\omega_0)~,\\
    \lim_{t \to \infty} \int_0^{\infty} \d \omega \,\frac{\sin[(\omega +\omega_0)t]}{\omega+ \omega_0} \phi(\omega) &= 0~,\\
    \lim_{t \to \infty} \int_0^{\infty}\d \omega\,\frac{\sin^2[(\omega \pm \omega_0)t/2]}{\omega\pm \omega_0}\phi(\omega)& =\frac{1}{2} \P\int_0^{\infty}\d \omega\,\frac{1}{\omega\pm\omega_0}\,\phi(\omega)~,
\end{align}
where $\phi$ is an arbitrary test function and $\P$ denotes the Cauchy principal value. We observe that, in the $t\to \infty$ limit, terms oscillating at sum frequency $\omega + \omega_0$ disappear from the equations for the Kossakowski matrix elements.
In this case, we have
\begin{align}
    \gamma_{11}^\infty & = 2\pi J(\omega_0)\bigl[n(\omega_0) +1 \bigr]~, \\
    \gamma_{22}^\infty & = 2\pi J(\omega_0)\,n(\omega_0) ~,\\
    \Re \gamma_{12}^{\infty} & = \alpha \pi J(\omega_0) \bigl[2n(\omega_0) +1 \bigr] ~,\\
    \Im \gamma_{12}^\infty & = -\alpha \P\int_{0}^\infty \d \omega\,J(\omega) \bigl[2n(\omega) +1 \bigr] \left[\frac{2\omega_0}{\omega^2- \omega_0^2}\right]~,
\end{align}
where $\gamma^\infty := \lim_{t\to\infty} \gamma(t)$. We immediately see that 
\begin{equation}
    \tr \gamma^\infty = 2\pi J(\omega_0)\coth \left(\frac{\beta \omega_0}{2}\right) \geq 0~.
\end{equation}
For $\alpha = 0$, we have
\begin{equation}
    \det \gamma^{\infty} = 4\pi^2 J^2(\omega_0)n(\omega_0)\bigl[n(\omega_0) +1 \bigr] \geq 0~,
\end{equation}
and thus $\gamma^\infty$ is always positive-semidefinite for $\alpha = 0$, as verified numerically beyond the weak-coupling limit. A necessary condition for the positive-semidefiniteness of $\gamma^\infty$ at finite $\alpha$ is derived as follows:
\begin{equation}
    \gamma_{11}^\infty \gamma_{22}^\infty \geq |\gamma_{12}^\infty|^2\geq \bigl[\Re\gamma_{12}^\infty\bigr]^2~,
\end{equation}
or 
\begin{equation}
    4  n(\omega_0)\bigl[n(\omega_0)+1\bigr] \geq \alpha^2 \bigl[2n(\omega_0)+1\bigr]^2~.
\end{equation}
Therefore, the condition
\begin{equation}
    \alpha > 2\frac{ \sqrt{ n(\omega_0)\bigl[1 + n(\omega_0)\bigr]}}{ 1+2n(\omega_0) } = \sech \left(\frac{\beta \omega_0}{2}\right)
\end{equation}
is sufficient for eternal non-Markovianity of the dynamics in the weak-coupling scenario.


%
\section{Low-temperature limit}
\label{app:lowT}
We want to investigate the low-temperature limit of the gQBM master equation~\eqref{sm-eq:gme}. We start by taking the $\beta \to \infty$ limit of the physical-time GF~\eqref{sm-eq:re-chi}-\eqref{sm-eq:im-chi}:
\begin{align}
\label{sm-eq:re-chi-lowt}
    \lim_{\beta \to \infty}\Re \chi(t) & = \frac{1}{4}\int_{0}^{\infty} \d \omega \,J(\omega)
    \begin{pmatrix}
        (1+\alpha)^2\cos(\omega t) & (1-\alpha^2)\sin(\omega t)\\[5pt]
         -(1-\alpha^2)\sin(\omega t) &  (1-\alpha)^2\cos(\omega t)
    \end{pmatrix}
    :=
    \frac{1}{2}
    \begin{pmatrix}
         (1+\alpha)^2\chi_c(t) & (1-\alpha^2)\chi_s(t)\\[5pt]
         -(1-\alpha^2)\chi_s(t) &  (1-\alpha)^2\chi_c(t)
    \end{pmatrix}
    ~,\\[8pt]
\label{sm-eq:im-chi-lowt}
    \lim_{\beta \to \infty} \Im \chi (t) & = -\frac{1}{4}\int_{0}^{\infty} \d \omega \,J(\omega)\begin{pmatrix}
         (1+\alpha)^2\sin(\omega t) &  -(1-\alpha^2)\cos(\omega t)\\[5pt]
          (1-\alpha^2)\cos(\omega t) &  (1-\alpha)^2\sin(\omega t)
    \end{pmatrix}
    :=
    -\frac{1}{2}
    \begin{pmatrix}
         (1+\alpha)^2\chi_s(t) & -(1-\alpha^2)\chi_c(t)\\[5pt]
         (1-\alpha^2)\chi_c(t) &  (1-\alpha)^2\chi_s(t)
    \end{pmatrix}~,
\end{align}
where we defined
\begin{align}
    \chi_c(t) & := \frac{1}{2} \int_{0}^\infty \d \omega \,J(\omega) \cos(\omega t)~, \\
    \chi_s(t) & := \frac{1}{2} \int_{0}^\infty \d \omega \,J(\omega) \sin(\omega t)~.
\end{align}
Now, to get an intuition about the phenomenology of the low-temperature limit, we perform a weak-coupling expansion in the constant $\eta$ defined in Eq.~\eqref{sm-eq:dos}. By definition, we have $\chi = \O(\eta)$, and from the Dyson equations~\eqref{eq:dyson-retarded}-\eqref{eq:dyson-keldysh} we see that
\begin{align}
    \G^R &= \C^R  + \O(\eta^2) = 2\Im\chi + \O(\eta^2)~, \\
    \G^K &= \C^K  + \O(\eta^2) = 2 \Re\chi + \O(\eta^2)~.
\end{align}
Thus, from Eq.~\eqref{sm-eq:gammaxx}-\eqref{sm-eq:imgammaxp} we see that
\begin{align}
    \gamma_{xx}(t) & = -\int_0^t \d \tau\Bigl[(1+\alpha)^2\chi_c(t-\tau)\cos(\tau-t) - (1-\alpha^2)\chi_s(t-\tau)\sin(\tau-t)\Bigr] + \O(\eta^2)\\[5pt]
    & :=  -(1+\alpha)^2 \Gamma_{cc}(t)+(1-\alpha)^2 \Gamma_{ss}(t) + \O(\eta^2)\notag  ~,\\[5pt]
    \gamma_{pp}(t) & = -\int_0^t \d \tau\Bigl[(1-\alpha)^2\chi_c(t-\tau)\cos(\tau-t) - (1+\alpha)^2\chi_s(t-\tau)\sin(\tau-t)\Bigr] + \O(\eta^2)\\[5pt]
    & =  -(1-\alpha)^2 \Gamma_{cc}(t)+(1+\alpha)^2 \Gamma_{ss}(t)+ \O(\eta^2) \notag ~,\\[5pt]
    \Re\gamma_{xp}(t) & = 2\alpha\int_{0}^t \d \tau\,\chi_c(t-\tau) \sin(\tau-t) + \O(\eta^2)\\[5pt]
    & := 2\alpha \Gamma_{cs}(t)+ \O(\eta^2) \notag~,\\[5pt]
    \Im \gamma_{xp}(t) & = i \int_{0}^t \d \tau\bigl[(1-\alpha^2) \chi_c(t-\tau) \cos(\tau-t) - (1+\alpha^2)\chi_s(t-\tau) \sin(\tau-t) \bigr] + \O(\eta^2)\\[5pt]
    & = i \bigl[(1-\alpha^2) \Gamma_{cc}(t)- (1+\alpha^2) \Gamma_{ss}(t)\bigr] + \O(\eta^2) \notag~.
\end{align}
Thus, we have
\begin{equation}
    \lim_{\beta \to \infty} \det\gamma(t) = -4\alpha^2 \bigl[\Gamma_{ss}^2(t) +\Gamma_{cs}^2(t) \bigr]- 4\alpha^2(3-\alpha^2)\Gamma_{cc}(t) \Gamma_{ss}(t) + \O(\eta^3)~.
\end{equation}
As we see from Fig.~\eqref{fig:3} of the main text, numerical convergence to the low temperature limit is already achieved for $\beta = 100$, as the eigenvalues of $\gamma(t)$ for $\beta = 50$ and $\beta = 100$ are identical within a relative tolerance of $0.5\%$. We verify numerically that, at $\beta = 100$, for $\alpha\neq 0$, we always have $\det \gamma(t)<0$, i.e. the model is eternally non-Markovian for any $\alpha \neq 0$. Conversely, in the case in which $\alpha = 0$ and for the same value of $\beta$, the Kossakowski matrix is
\begin{equation}
\label{sm-eq:lowT-gamma}
    \lim_{\alpha \to 0}\lim_{\beta \to \infty} \gamma(t) = 
    \begin{pmatrix}
        \gamma_{xx}(t) & -i \gamma_{xx}(t)\\
        i \gamma_{xx}(t) & \gamma_{xx}(t) 
    \end{pmatrix}
    + \O(\eta^2)~.
\end{equation}
The latter is rank $1$ at leading order in $\eta$, and therefore we have $\lim_{\beta \to \infty}\det \gamma(t) = \O(\eta^3)$. For the range of parameters explored in the present work, truncation at first order in $\eta$ is numerically verified to be exact within a relative tolerance of $<4\%$ on the time-integrated entries of the Kossakowski matrix, and we have $\det \gamma(t) < 10^{-4}$ at $\beta = 100$. Diagonalization of the Kossakowski matrix~\eqref{sm-eq:lowT-gamma} yields $\gamma_-(t) = \O(\eta^2)$ and $\gamma_{+}(t) = 2 \gamma_{xx}(t) + \O(\eta^2)$. Expressing the low-temperature limit of the gQBM master equation~\eqref{eq:quasi-gksl} in terms of these canonical rates yields Eq.~\eqref{eq:lowT-master} of the main text.
\section{Measures of information backflow for Gaussian states}
\setcounter{figure}{0}
\label{app:measures}
In this Appendix, we discuss how to characterize the dynamics of Gaussian states which evolution is governed by the GME~\eqref{sm-eq:gme} by writing ODEs for their first and second statistical moments. Then, we give some details about the Breuer--Laine--Piilo~\cite{BLPpaper} and entanglement revival~\cite{rivas2010entanglement} measures of non-Markovianity, and we provide formulas for the Umegaki relative entropy and the logarithmic negativity of entanglement for Gaussian states.
\subsection{Dynamics of continuous-variable Gaussian states}
The most general $n$-mode continuous-variable Gaussian state $\rho$ is uniquely identified by its displacement vector $\bar{\bm r}$ and covariance matrix (CM) $\sigma$~\cite{serafini2017quantum}, defined as
\begin{align}
    \bar{\bm r} &:= \tr\bigl[ {\bm r}\rho\bigr]~, \\
    \sigma_{ij} & := \tr\bigl[\bigl\{({r}_i - \bar{r}_i), ({r}_j-\bar{r}_j) \bigr\} \rho \bigr]~.
\end{align}
Here and in the following, ${\bm r} := ({\bm x}, {\bm p})^T$ is the quadratures vector. We can present such a Gaussian state as
\begin{equation}
    \rho = \frac{1}{\Z} \exp\left[-\frac{1}{2}({r}_i - \bar{r}_i) H_{ij}({r}_j - \bar{r}_j)\right]~,
\end{equation}
where 
\begin{equation}
    \Z = \sqrt{\det \bigl[\bigr(\sigma + i \Omega\bigr)/2\bigr]}
\end{equation}
and $H$ is positive-definite, real Hamiltonian matrix; $\Omega := \bigotimes_{i=1}^n\sigma_x$ is the standard symplectic form in dimension $2n$. $H$ and $\sigma$ are related by
\begin{align}
\label{eq:sigma-and-H}
    \sigma & = i \coth \left(\frac{i  H\Omega}{2} \right) \Omega~,\\
\label{eq:Handsigma}
    H &= 2i \Omega\operatorname{arcoth} \left(i \sigma \Omega\right)~,
\end{align}
where
\begin{align}
    \coth(x) &= \frac{e^x + e^{-x}}{e^x - e^{-x}} ~,\\
    \arcoth(x) &= \frac{1}{2}\log\left(\frac{x+1}{x-1}\right)~. 
\end{align}
In order for the state to be well-defined, $\sigma$ must satisfy the Robertson-Schr\"odinger uncertainty relation
\begin{equation}
    \sigma + i   \Omega \geq 0~.
\end{equation}
The evolution of the first and second moments of a Gaussian state $\rho$ under the master equation~\eqref{eq:gme} are defined by the following ODEs:
\begin{align}
\label{eq:ode-for-r}
    \frac{\d \bar{\bm r}}{\d t} &= \J \bar{\bm r}~,\\
\label{eq:ode-for-sigma}
    \frac{\d \sigma}{\d t} & = \J \sigma + \sigma \J^T + D~,
\end{align}
where we defined the drift and diffusion matrices
\begin{align}
\label{eq:Jmatrix}
    \J &:= 2\Omega\Im [\M]+\Omega h~,\\
\label{eq:Dmatrix}
    D & := -2 \Omega \Re\bigl[\M + \M^T\bigr] \Omega~,
\end{align}
where $\M$ is defined in Eq.~\eqref{sm-eq:M-matrix} and $h$ is defined in Eq.~\eqref{sm-eq:hmatrix}.

\subsection{Umegaki relative entropy between Gaussian states and BLP measure}
The Umegaki relative entropy is defined as
\begin{equation}
    D[\rho\|\tau] := \tr \bigl[\rho \bigl(\log \rho - \log \tau\bigr)\bigr]~.
\end{equation}
If $\rho$ and $\tau$ are bosonic Gaussian states with covariances $\sigma_\rho$ and $\sigma_\tau$ and displacement vectors $\bar {\bm r}_\rho$, $\bar {\bm r}_\tau$, the Umegaki is given by~\cite{Seshadreesan2018}
\begin{equation}
\label{eq:umegaki-long}
    D[\rho \| \tau] = \frac{1}{2} \left[ \log \left(\frac{\det \bigl[\sigma_\tau + i \Omega\bigr]}{\det \bigl[\sigma_\rho + i \Omega\bigr]}\right) + \frac{1}{2} \tr \bigl[\sigma_\rho \bigl(H_\tau - H_\rho\bigr)\bigr] + (\bar{\bm r}_\tau - \bar {\bm r}_\rho)^T H_\tau  (\bar{\bm r}_\tau - \bar {\bm r}_\rho)\right]~.
\end{equation}
In order to calculate the Umegaki, we need to compute the Hamiltonian via relation~\eqref{eq:Handsigma}. This is achieved by calculating the symplectic eigenvalues of $\sigma$. In the single-mode scenario, the only symplectic eigenvalue of $\sigma$ is $\nu = \sqrt{\det \sigma}$~\cite{olivares2012quantum}. Let us consider the matrix $M :=i\sigma \Omega$. The Cayley-Hamilton theorem states that $M$ satisfies the characteristic equation
\begin{equation}
    M^2 - \tr[M] M + \det[M] \1 = 0~.
\end{equation}
Since $\tr[M] = 0$, we have $M^2 = - \det[M] \1$, and since $\det M = -\det \sigma := -\nu^2$ we have $M^2 = \nu^2\1$. Therefore,
\begin{equation}
    H = 2i \Omega \arcoth \left(i \sigma \Omega\right) = 2i \Omega\arcoth \left(\nu\right) \frac{i\sigma \Omega}{\nu} = -\frac{2}{ \nu} \arcoth \left(\nu\right) \Omega\sigma\Omega~.
\end{equation}
Finally, we observe that $-\Omega \sigma \Omega = \det[\sigma] \sigma^{-1} = \nu^2 \sigma^{-1}$. Therefore,
\begin{equation}
\label{eq:hamiltonian-matrix}
    H =2\nu \arcoth \left(\nu\right) \sigma^{-1} = \nu \ln \left(\frac{\nu +1}{\nu-1}\right) \sigma^{-1}~.
\end{equation}
In the simple case considered in this work, in which $\sigma_\rho \equiv \sigma_\tau$, the equation for $S$ is simply
\begin{equation}
    \label{eq:umegaki-simple}
    \begin{split}
        D\bigl[\rho \|\tau\bigr] 
        & = \frac{1}{2} (\bar{\bm r}_\tau - \bar {\bm r}_\rho)^T H_\tau  (\bar{\bm r}_\tau - \bar {\bm r}_\rho)\\
        & = \nu_\tau \arcoth(\nu_\tau) (\bar{\bm r}_\tau - \bar {\bm r}_\rho)^T \sigma_\tau^{-1}  (\bar{\bm r}_\tau - \bar {\bm r}_\rho)\\
        & = \nu_\rho \arcoth(\nu_\rho) (\bar{\bm r}_\tau - \bar {\bm r}_\rho)^T \sigma_\rho^{-1}  (\bar{\bm r}_\tau - \bar {\bm r}_\rho)~.
    \end{split}
\end{equation}
Employing Eq.~\eqref{eq:umegaki-simple} together with the ODEs~\eqref{eq:ode-for-r}-\eqref{eq:ode-for-sigma}, we are able to evaluate numerically the Umegaki relative entropy $D\bigl[ \rho(t) \| \sigma(t)\bigr]$ as a function of time $t$ as shown in Fig.~\ref{fig:4}b of the main text.
Given a grid of $N$ points in the interval $[0, T]$ with grid size $\Delta t$ s.t. $t_l = l\Delta t$, we can approximate the BLP measure $\N$ [Eq.~\eqref{eq:N} in the main text] as
\begin{equation}
    \N\approx {\sum_{l=1}^{N-1}}{\phantom {\big|}}^{\!\!\!*} \Bigl(D\bigl[\rho(t_{l+1})\| \tau(t_{l+1})\bigr]- D\bigl[\rho(t_{l})\| \tau(t_{l})\bigr]\Bigr)~,
\end{equation}
where the starred sum is restricted to $l$ s.t. $D\bigl[\rho(t_{l+1})\| \tau(t_{l+1})\bigr]- D\bigl[\rho(t_{l})\| \tau(t_{l})\bigr] >0$. Initial states $\rho(0)$ and $\tau(0)$ are differently-displaced quasi-coherent states with $\bar {\bm r}_\rho = (3, -3)^T$, $\bar {\bm r}_\tau = (-3, 3)^T$ and $\sigma_{\rho} = \sigma_\tau = (1+\epsilon) \1_{2\times2}$, with $\epsilon = 0.001$.

\begin{figure}
    \centering
    \includegraphics[width=0.5\linewidth]{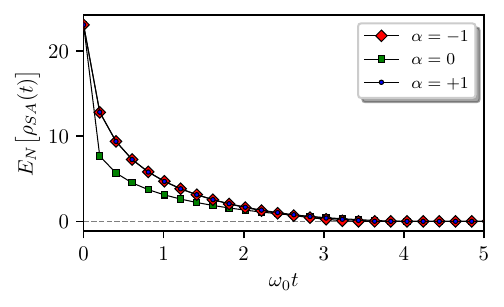}
    \caption{Entanglement negativity as a function of time for different values of the control parameter $\alpha$. The initial state is a two-mode vacuum squeezed state defined by Eq.~\eqref{sm-eq:2modesqueezed} with squeezing $r=8$. Other parameters are the same specified above in App.~\ref{app:gme}.}
    \label{smfig:4}
\end{figure}

\subsection{Logarithmic negativity of two-mode Gaussian states}
Another popular criterion for non-Markovianity is entanglement revival~\cite{rivas2010entanglement}. Given some initial maximally-entangled system-ancilla state $\rho_{SA}(0)$, and some entanglement measure $E$, we can introduce 
\begin{equation}
\label{sm-eq:entanglement-witness}
    \I^{(E)} := \int_{0}^T \d t \,\bigl|\dot E[\rho_{SA}(t)]\bigr| - \Delta E~,
\end{equation}
where $\Delta E := E[\rho_{SA}(T)] - E[\rho_{SA}(0)]$, such that $\I^{(E)} >0$ is a sufficient condition for non-Markovianity. As an entanglement measure, we employ the logarithmic negativity
\begin{equation}
    E_N[\rho_{SA}]:= \log_2 \|\bar \rho_{SA}\|_1~,
\end{equation}
where $\bar \rho$ denotes the partial transpose of a bipartite quantum state $\rho$, and $\|\bullet\|_1$ is the trace norm. The logarithmic negativity finds a particularly simple expression for the case of a two-mode Gaussian state $\rho_{G, 1+1}$, identified by its covariance matrix
\begin{equation}
    \sigma = 
    \begin{pmatrix}
        \sigma_{SS} & \sigma_{SA}\\[3pt]
        \sigma_{SA}^T & \sigma_{AA}
    \end{pmatrix}~,
\end{equation}
The logarithmic negativity in this case is~\cite{serafini2017quantum}
\begin{equation}
\label{eq:logarithmic-negtivity}
    E_N[\rho_{G, 1+1}] = \max \bigl\{0, -\log_2 \bar \nu_-\bigr\}~,
\end{equation}
where $\bar \nu_-$ is the lesser of the two symplectic eigenvalues of $\bar \rho_{G,1+1}$, given by
\begin{equation}
    \bar \nu_{-}^2 = \frac{1}{2}\Bigl(\bar \Delta  - \sqrt{\bar \Delta^2- 4 \det \sigma}\Bigr) ~,
\end{equation}
where
\begin{equation}
    \bar \Delta = \det \sigma_{SS} + \det \sigma_{AA} - 2\det \sigma_{SA}
\end{equation}
is the Seralian of $\bar \rho_{G, 1+1}$ (as a function of the covariances of $ \rho_{G, 1+1}$).

To calculate the witness in Eq.~\eqref{sm-eq:entanglement-witness}, following~\cite{rivas2010entanglement}, we initialize the system-ancilla composite in the two-mode vacuum squeezed state $\rho_{SA}^{(r)}$ identified by its covariance matrix
\begin{equation}
\label{sm-eq:2modesqueezed}
    \sigma_r = 
    \begin{pmatrix}
        \cosh(2r) \1 & \sinh(2r) \sigma_z\\
        \sinh(2r) \sigma_z &\cosh(2r)\1
    \end{pmatrix}~,
\end{equation}
which approaches the maximally entangled state as $r \to \infty$. 

We evaluate numerically the logarithmic negativity of entanglement for $r \in [0,8]$. Remarkably, as shown in Fig.~\ref{smfig:4}, this quantity is always decreasing independently of the value of the control parameter $\alpha$, and it is thus blind to non-Markovianity within the model and parameters explored in this work.

\fi

\end{document}